\documentclass[%
 aip,
 amsmath,amssymb,
reprint,
]{revtex4-1}

\usepackage{graphicx}
\usepackage{dcolumn}
\usepackage{bm}

\usepackage[utf8]{inputenc}
\usepackage[T1]{fontenc}
\usepackage{mathptmx}
\usepackage{etoolbox}
\usepackage{amsmath}
\usepackage{physics}
\usepackage{siunitx}
\usepackage[dvipsnames]{xcolor}
\usepackage{comment}
\usepackage{dsfont}
\usepackage{pifont}

\newcommand{\super}[1]{\mathcal{#1}}

\newcommand{\bigcdot}{\boldsymbol{\cdot}}

\makeatletter
\def\@email#1#2{%
 \endgroup
 \patchcmd{\titleblock@produce}
  {\frontmatter@RRAPformat}
  {\frontmatter@RRAPformat{\produce@RRAP{*#1\href{mailto:#2}{#2}}}\frontmatter@RRAPformat}
  {}{}
}%
\makeatother
\begin{document}

\preprint{AIP/123-QED}

\title[]{Influence of excitonic coupling, static disorder, and coherent dynamics in action-2D electronic spectroscopy of a molecular dimer model}

\author{Matteo Bruschi}
\affiliation{Dipartimento di Scienze Chimiche, Università degli Studi di Padova, via Marzolo 1, Padua 35131, Italy.}
\email{matteo.bruschi@unipd.it}
\author{Roberto Zambon}
\affiliation{Dipartimento di Scienze Chimiche, Università degli Studi di Padova, via Marzolo 1, Padua 35131, Italy.}
\author{Federico Gallina}
\affiliation{Dipartimento di Scienze Chimiche, Università degli Studi di Padova, via Marzolo 1, Padua 35131, Italy.}
\author{Barbara Fresch}
\affiliation{Dipartimento di Scienze Chimiche, Università degli Studi di Padova, via Marzolo 1, Padua 35131, Italy.}
\affiliation{Padua Quantum Technologies Research Center, Università degli Studi di Padova, via Gradenigo 6/A, Padua 35131, Italy.}
\email{barbara.fresch@unipd.it}
\date{\today}

\begin{abstract}
We investigate the spectral features of Action-2D Electronic Spectroscopy (A-2DES) in a molecular dimer model across different regimes of excitonic coupling. 
By explicitly including a second-excited state for each chromophore, we simulate A-2DES spectra ranging from the non-interacting limit to the strong-coupling case, focusing on the significance of cross peaks. 
While for weak excitonic coupling, cross peaks can be understood as the incoherent mixing of linear signals of the two chromophores, these features reflect excitonic delocalization as the coupling increases.
We highlight that A-2DES offers enhanced sensitivity to coherent excited-state dynamics, particularly in the intermediate-coupling regime, where it provides higher contrast compared to its coherent-detected counterpart.
Finally, we show that static disorder reduces the relative amplitude of cross peaks compared to diagonal features in a way that depends on the excitonic coupling. 
Notably, the relative suppression of cross peaks decreases with the strength of the excitonic coupling, implying that spectral features related to incoherent mixing are less prominent in inhomogeneous samples.
These findings support the potential of A-2DES for investigating excitonic dynamics in small multi-chromophoric systems.
\end{abstract}

\maketitle

\section{Introduction}

Action-2D Electronic Spectroscopy (A-2DES) represents a significant addition to the suite of non-linear ultrafast spectroscopic techniques as introduces both technical and conceptual novelties compared to its coherent-detected counterpart (C-2DES).\cite{tiwari:2021, karki:2022} In A-2DES, a sequence of laser pulses arranged in a fully collinear geometry excites the system, resulting in the emission of an incoherent signal which reflects the distribution of excited-state populations. The relevant components of the optical response can then be selectively isolated using phase-cycling\cite{tian:2003, tan:2008} or phase-modulation\cite{tekavec:2007} schemes.
Depending on the nature of the sample, different types of incoherent signals can be detected, e.g., fluorescence,\cite{lott:2011, tiwari:2018a, karki:2019} photocurrent,\cite{nardin:2013, karki:2014, bolzonello:2021} photoions\cite{roeding:2018} and photoelectrons\cite{uhl:2021}, endowing the technique with a unique function-oriented response.\cite{bakulin:2016}
Furthermore, when combined with microscopy \cite{tiwari:2018b} and single-molecule \cite{fersch:2023, jana:2024} methods, it allows for achieving spatial resolution below the diffraction limit.
To date, A-2DES has been applied to investigate the optical response of a variety of different systems, such as atomic vapors,\cite{tekavec:2007, bruder:2015, bruder:2019} organic dyes,\cite{mueller:2018, mueller:2020} molecular dimers\cite{lott:2011, tiwari:2018a, maly:2020} and aggregates,\cite{meza:2021} light-harvesting complexes,\cite{tiwari:2018b, karki:2019, lopez_ortiz:2024, javed:2024} organic \cite{bian:2020, bolzonello:2021, casotto:2024} and inorganic \cite{karki:2014, mueller:2021, zhou:2020} solar cells, plasmon-\cite{pres:2023} and exciton-polaritons.\cite{autry:2020}

From these studies, it appears clearly that the origin and interpretation of the spectral features in A-2DES are not necessarily related to those of C-2DES, although the two techniques ultimately probe the same underlying ultrafast dynamics.\cite{maly:2020b, sun:2024}
While in C-2DES, cross peaks at early waiting times exclusively indicate excitonic delocalization, in A-2DES, they may also result from Exciton-Exciton Annihilation (EEA) occurring during signal emission.\cite{pedromoortiz:2012, maly:2018, schroter:2018, kunsel:2019, kuhn:2020, maly:2020} 
For weakly-interacting systems, the emergence of such cross peaks can be equivalently interpreted within the framework of incoherent mixing.\cite{gregoire:2017, kalaee:2019, bruschi:2023} 
In this case, cross peaks can be understood as given by the product of linear signals of independent chromophores.\cite{gregoire:2017, bruschi:2023}
In multi-chromophoric systems, the contribution of incoherent mixing can dominate the A-2DES spectrum, potentially hiding excited-state dynamics.\cite{bolzonello:2023, javed:2024, lopez_ortiz:2024}
In contrast, in strongly-interacting systems, it is no longer valid to consider chromophores as independent entities, since excitonic coupling induces excitonic delocalization, energy splitting, and transition dipole moment redistribution between the states.\cite{maly:2018, schroter:2018, kunsel:2019, maly:2020}
In this case, cross peaks constitute an essential part of the non-linear optical response of the coupled system.

The effect of the excitonic coupling on the amplitude of the spectral features at early-waiting time has been previously analyzed in Refs. \onlinecite{maly:2018, schroter:2018, kunsel:2019, maly:2020}.
In this work, we revisit this issue to clarify how the nature of spectral features in A-2DES changes from the non-interacting limit to the strong-coupling case. To this end, we consider a molecular dimer model, explicitly accounting for the delocalization between two-exciton and second-excited states induced by excitonic coupling. This represents a minimal model for establishing a relationship between the EEA process and excitonic delocalization. In general, EEA is described as an incoherent mechanism resulting from the stochastic encounter of two excitons in the system, which is typically modeled phenomenologically as a bimolecular process.\cite{suna:1970, valkunas:1995}
Beyond this description, there have been few attempts to describe EEA in a coherent framework.\cite{renger:1997, ryzhov:2001, bruggemann:2002}
In this framework, the EEA process results from excitonic delocalization between the second-excited states of individual chromophores and two-exciton states on different chromophores. Due to delocalization, excitonic states inherit the fast internal conversion character of second-excited states, making relaxation to the one-exciton manifold an important decay channel. Notably, this description has been recently used to reveal other coherent effects in the EEA process.\cite{tempelaar:2017, suss:2020a, suss:2020b, kumar:2023, maly:2023}

Equipped with a mechanistic model of EEA, we simulate the optical response of a molecular dimer as a function of the excitonic coupling, for both A-2DES and C-2DES.
We begin by analyzing the change in the spectral features at early-waiting times, focusing on the interplay between excitonic delocalization and EEA in shaping the spectra (Sec. \ref{sec:IIIA}).
In the limit of non-interacting chromophores, cross peaks are absent regardless of the detection method. 
In the strong-coupling case, cross peaks observed in A-2DES probe the same excitonic states as in C-2DES, though their amplitude pattern differs due to the distinct contribution from the two-exciton manifold.
Between these two limits, the prominence of cross peaks varies with the strength of the excitonic coupling.
Based on their amplitude in A-2DES, we propose a classification of different excitonic coupling regimes.
Notably, for intermediate coupling, A-2DES exhibits pronounced cross peaks that reveal excited-state dynamics with higher contrast compared to C-2DES, where cross peaks are barely distinguishable from the background.
In our model, we observe oscillatory dynamics during the waiting time related to coherences between one-exciton eigenstates (Sec. \ref{sec:IIIB}). However, we believe that A-2DES enhances the detectability of coherent dynamics regardless of its origin, i.e., electronic, vibronic, or vibrational.
Another important finding of our analysis is the positive role of inhomogeneous broadening in attenuating incoherent mixing features in A-2DES spectra (Sec. \ref{sec:IIIC}).
By simulating A-2DES spectra with varying degrees of static disorder and interpreting the results using the stochastic lineshape theory, we conclude that cross peaks lack or exhibit rephasing capability depending on the excitonic coupling. 
On the one hand, this implies that the relative amplitude of cross peaks related to incoherent mixing is reduced in disordered systems. 
On the other hand, cross peaks become more robust to disorder when the excitonic coupling is significant, that is when they carry actual non-linear information.

\section{Theory and Simulations}
\label{sec:theory}

\subsection{Model system -- Molecular dimer}
\label{sec:system}

\begin{figure}
    \centering
    \includegraphics[width=\columnwidth]{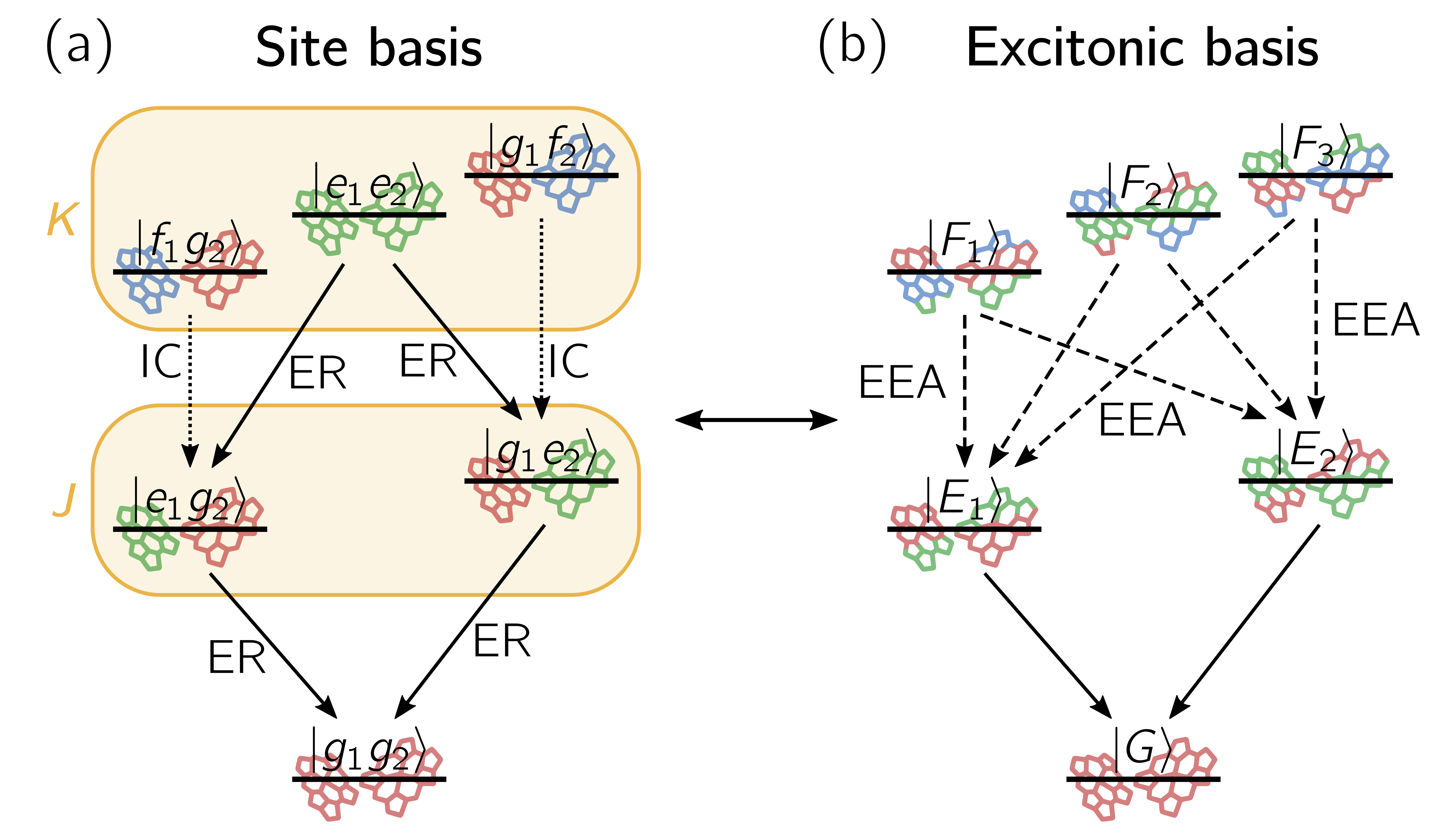}
    \caption{Model of a molecular dimer, where each chromophore is described as a three-level electronic system. The states can be represented in the (a) localized site basis and (b) delocalized excitonic basis. The yellow boxes indicate the states of the one- and two-exciton manifolds interacting via excitonic coupling $J$ and $K$, respectively. The arrows represent the processes of Exciton Recombination (ER), Internal Conversion (IC), and Exciton-Exciton Annihilation (EEA).}
    \label{fig:dimer_scheme}
\end{figure}

In the following, we consider a molecular dimer in which each monomer is treated as a three-level electronic system, consisting of a ground state $\ket{g_{n}}$, a first-excited state $\ket{e_{n}}$ and a second-excited state $\ket{f_{n}}$ (for $n=1, 2$). The Hamiltonian of the $n$-th chromophore is:
\begin{equation}
    \hat{H}_{n} = \epsilon_{g_n} \ketbra{g_{n}}{g_{n}} + \epsilon_{e_n} \ketbra{e_{n}}{e_{n}} + \epsilon_{f_n} \ketbra{f_{n}}{f_{n}}
\end{equation}
where $\epsilon_{g_n}$, $\epsilon_{e_n}$ and $\epsilon_{f_n}$ are the energies of the ground, first-excited and second-excited states, respectively.

Depending on the number of excitons, the states of the molecular dimer can be categorized into different manifolds. 
In the (localized) site basis (Fig. \ref{fig:dimer_scheme}a), these manifolds are: the common ground state, where both molecules are in the ground state (i.e., $\ket{g_{1}g_{2}}$); the one-exciton manifold, where a molecule is in the first-excited state and the other is in the ground state (i.e., $\ket{e_{1}g_{2}}$, $\ket{g_{1}e_{2}}$); and the two-exciton manifold, where both molecules are in the first-excited state (i.e., $\ket{e_{1}e_{2}}$) or a molecule is in the second-excited state and the other is in the ground state (i.e., $\ket{f_{1}g_{2}}$, $\ket{g_{1}f_{2}}$). Note that states beyond the two-exciton manifold are ignored in this description because they will not be populated at the fourth order in the light-matter interaction. 

Assuming negligible coupling between states belonging to different manifolds, the Hamiltonian of the molecular dimer is block diagonal:
\begin{equation}
    \hat{H} = \hat{H}^{(0)} + \hat{H}^{(1)} + \hat{H}^{(2)}
    \label{eq:system_Hamiltonian}
\end{equation}
with the ground-state Hamiltonian:
\begin{equation}
    \hat{H}^{(0)} = \epsilon_{g_{1}g_{2}} \ketbra{g_{1}g_{2}}{g_{1}g_{2}}
\end{equation}
the one-exciton Hamiltonian:
\begin{equation}
\begin{split}
    \hat{H}^{(1)} =& \epsilon_{e_{1}g_{2}} \ketbra{e_{1}g_{2}}{e_{1}g_{2}} + \epsilon_{g_{1}e_{2}} \ketbra{g_{1}e_{2}}{g_{1}e_{2}} \\
    & + J \left( \ketbra{e_{1}g_{2}}{g_{1}e_{2}} +  h.c. \right)
\end{split}
\end{equation}
and the two-exciton Hamiltonian:
\begin{equation}
\begin{split}    
    \hat{H}^{(2)} =& \epsilon_{e_{1}e_{2}} \ketbra{e_{1}e_{2}}{e_{1}e_{2}} \\
    &+ \epsilon_{f_{1}g_{2}} \ketbra{f_{1}g_{2}}{f_{1}g_{2}} + \epsilon_{g_{1}f_{2}} \ketbra{g_{1}f_{2}}{g_{1}f_{2}} \\
    &+ K \left(\ketbra{e_{1}e_{2}}{f_{1}g_{2}} + \ketbra{e_{1}e_{2}}{g_{1}f_{2}} + h.c. \right)
\end{split}
\end{equation}
where $\epsilon_{a_{1}b_{2}} = \epsilon_{a_{1}} + \epsilon_{b_{2}}$ (for $a,b = g,e,f$) is the energy of the state $\ket{a_{1}b_{2}}$, while $J$ and $K$ are the excitonic couplings between states in the one- and two-exciton manifold, respectively. In the following, we assume these couplings to be proportional $K = \alpha J$, where $\alpha$ is a scaling factor.\cite{maly:2020} Note that the direct coupling between the second-excited states of the two chromophores has been neglected.

Due to the structure of the Hamiltonian, each block can be diagonalized independently to obtain the eigenstates of the system, which are (delocalized) excitonic states given by the linear combination of (localized) site states (Fig. \ref{fig:dimer_scheme}b).\cite{maly:2020} For the ground-state manifold, the eigenstate is readily available:
\begin{equation}
        \ket{G} = \ket{g_{1}g_{2}}
\end{equation}
while for the one-exciton manifold are:
\begin{equation}
    \ket{E_{k}} = a_{1k} \ket{e_{1} g_{2}} + a_{2k} \ket{g_{1} e_{2}}
    \label{eq:one_exciton_states}
\end{equation}
with coefficients $a_{nk}$ (for $k = 1, 2$), and for the two-exciton manifold are:
\begin{equation}
    \ket{F_{l}} = b_{1l} \ket{e_{1} e_{2}} + b_{2l} \ket{f_{1} g_{2}} + b_{3l} \ket{g_{1} f_{2}}.
\label{eq:two_exciton_states}
\end{equation}
with coefficients $b_{ml}$ (for $m,l = 1, 2, 3$). The eigenstates and eigenvalues are explicitly reported in the Supplementary Material as a function of the site energies and the excitonic coupling.
In Table \ref{tab:tableI}, we report the parameters of the molecular dimer model used in the simulations.

\begin{table}
\caption{\label{tab:tableI}Parameters used in the simulations.}
\begin{ruledtabular}
\begin{tabular}{l c l c}
Parameter & Value & Parameter & Value \\
\hline
$\epsilon_{g_{1}}$ & $0.00$ eV & $\epsilon_{g_{2}}$ & $0.00$ eV \\
$\epsilon_{e_{1}}$ & $1.46$ eV & $\epsilon_{e_{2}}$ & $1.55$ eV\\
$\epsilon_{f_{1}} = 2 \epsilon_{e_{1}}$ & $2.92$ eV & $\epsilon_{f_{2}} = 2 \epsilon_{e_{1}}$ & $3.10$ eV  \\
$J$ & $[-45,45]$ meV &  $\alpha$ & $0.5$ \\
$\mu_{e_{n}g_{n}} E^{0}_{i}$ & 1 eV & $\beta$ & 0 \\
$t_{1}$ & $[0, 300]\ \si{fs}$ & $k_{\text{ER}}/h$ & $(10\ \si{ns})^{-1}$ \\ 
$t_{2}$ & $[0, 100]\ \si{fs}$ & $k_{\text{IC}}/h$ & $(1\ \si{ps})^{-1}$ \\
$t_{3}$ & $[0,300]\ \si{fs}$ & $k_{\text{D}}/h$ & $(100\ \si{fs})^{-1}$ \\
\end{tabular}
\end{ruledtabular}
\end{table}

\subsection{Quantum dynamics -- Lindblad quantum master equation}
\label{sec:dynamics}

The system dynamics is described using the Lindblad quantum master equation:\cite{breuer:2002}
\begin{equation}
    \frac{d}{dt}\rho(t) = \super{L} \rho(t)
\label{eq:lindblad_eq}
\end{equation}
with the system density matrix $\rho(t)$ and the Liouvillian superoperator:
\begin{equation}
    \super{L}\bigcdot = -\frac{i}{\hbar} \left[\hat{H}, \bigcdot \right] + \sum_{j} \frac{k_{j}}{\hbar} \left( \hat{L}_{j} \bigcdot \hat{L}_{j}^{\dagger} - \frac{1}{2} \left\{ \hat{L}_{j}^{\dagger} \hat{L}_{j}, \bigcdot \right\} \right)
\end{equation}
where $[\bigcdot, \bigcdot]$ and $\{\bigcdot, \bigcdot \}$ are the commutator and anti-commutator, respectively. 
The effect of Lindblad operators $\hat{L}_{j}$ on the system density matrix is to induce decoherence and relaxation processes due to the interaction with the environment with associated rates $k_{j}$.

In the following, we assume that each chromophore interacts with a local independent environment.
In the site basis, relaxation between states of different manifolds is due to Exciton Recombination (ER) and Internal Conversion (IC) processes (Fig. \ref{fig:dimer_scheme}a), associated with $\hat{L}_{g_{n} e_{n}} = \ketbra{g_{n}}{e_{n}}$ and $\hat{L}_{e_{n} f_{n}} = \ketbra{e_{n}}{f_{n}}$, respectively.
Furthermore, we consider pure-dephasing processes, $\hat{L}_{e_{n}e_{n}} = \ketbra{e_{n}}{e_{n}}$ and $\hat{L}_{f_{n}f_{n}} = \ketbra{f_{n}}{f_{n}}$, which represent fluctuations of the site energies and cause transitions within each manifold.

In the site basis, EEA can be visualized as a two-step process (Fig. \ref{fig:dimer_scheme}a): first, energy transfer from the two-exciton state (i.e., $\ket{e_{1}e_{2}}$) to second-excited states (i.e., $\ket{f_{1}g_{2}}$ and $\ket{g_{1}f_{2}}$) occurs as a result of the excitonic coupling $K$; then, the rapid IC to the one-exciton states (i.e., $\ket{e_{1}g_{2}}$ and $\ket{g_{1}e_{2}}$) leads to the net loss of one excitation.
Since we want to investigate the spectral features in A-2DES ranging from zero to strong excitonic coupling, we use an eigenstate picture of the EEA process. In this case, EEA reflects the delocalization of states within the two-exciton manifold (Fig. \ref{fig:dimer_scheme}b): the two-exciton state (i.e., $\ket{e_{1}e_{2}}$) mixes with second-excited states (i.e., $\ket{f_{1}g_{2}}$ and $\ket{g_{1}f_{2}}$) inheriting their rapid IC character.
Indeed, due to delocalization, the eigenstates may have different relaxation rates compared to the site states.\cite{bruggemann:2002} As reported in the Supplementary Material, the EEA rate increases rapidly as soon as the participation ratio of the eigenstates in the two-exciton manifold deviates from unity.
In the excitonic basis, the relaxation rate from a two-exciton state to a one-exciton state is:
\begin{equation}
\begin{split}
    k_{E_k F_l} =& k_{g_1 e_1} \abs{a_{2k}}^2 \abs{b_{1l}}^2 + k_{g_2 e_2} \abs{a_{1k}}^2 \abs{b_{1l}}^2 \\
    &+ k_{e_1 f_1} \abs{a_{1k}}^2 \abs{b_{2l}}^2 + k_{e_2 f_2} \abs{a_{2k}}^2 \abs{b_{3l}}^2 
\end{split}
\label{eq:relaxation_rate_2}
\end{equation}
while that from a one-exciton state to the ground state is:
\begin{equation}
    k_{G E_k} = k_{g_1 e_1} \abs{a_{1k}}^2 + k_{g_2 e_2} \abs{a_{2k}}^2.
\label{eq:relaxation_rate_1}
\end{equation}
The values of the rates of exciton recombination $k_{\text{ER}} = k_{g_{n} e_{n}}$, internal conversion $k_{\text{IC}} = k_{e_{n} f_{n}}$, and pure-dephasing $k_{\text{D}} = k_{e_{n}e_{n}} = k_{f_{n}f_{n}}$ used in the simulations are reported in Table \ref{tab:tableI}.

\subsection{Light-Matter Interaction -- Action-2D electronic spectroscopy}
\label{sec:light}

\begin{figure}
    \centering
    \includegraphics[width=\columnwidth]{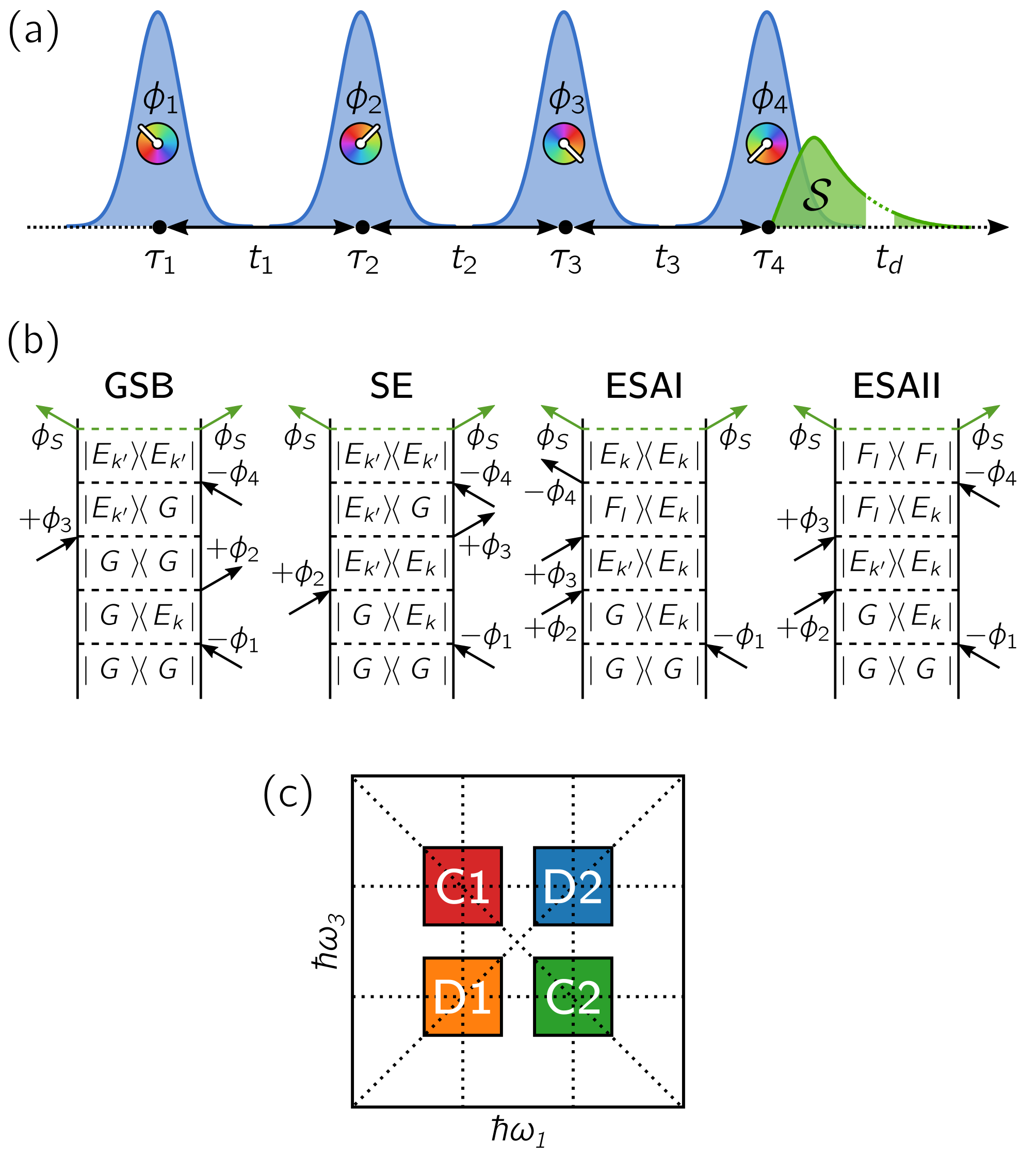}
    \caption{(a) Sequence of four laser pulses used in A-2DES experiments and emission of the incoherent signal. (b) Feynman diagrams for the considered system relative to Ground-State Bleaching (GSB), Stimulated Emission (SE), Excited-State Absorption I (ESAI), and Excited-State Absorption II (ESAII) pathways. (c) Position of the spectral features corresponding to the diagonal peaks (D1 and D2) and cross peaks (C1 and C2).}
    \label{fig:pulse_FD}
\end{figure}

As a result of the light-matter interaction, transitions between states of different manifolds are induced.
In the dipole approximation, the light-matter interaction Hamiltonian is:
\begin{equation}
    \hat{H}'(t) = - \hat{\mu} \cdot E(t)
\label{eq:light_matter_Hamiltonian}
\end{equation}
where $\hat{\mu}$ is the dipole moment operator and $E(t)$ is the electric field.
In terms of the site basis, the dipole moment operator of the dimer reads:
\begin{equation}
\begin{split}
    \hat{\mu} =& \mu_{e_{1}g_{1}} (\ketbra{e_{1}g_{2}}{g_{1}g_{2}} + \ketbra{e_{1}e_{2}}{g_{1}e_{2}} + h.c.) \\
    &+ \mu_{e_{2}g_{2}} (\ketbra{g_{1}e_{2}}{g_{1}g_{2}} + \ketbra{e_{1}e_{2}}{e_{1}g_{2}} + h.c.) \\
    &+ \mu_{f_{1}e_{1}} (\ketbra{f_{1}g_{2}}{e_{1}g_{2}} + h.c.) + \mu_{f_{2}e_{2}} (\ketbra{g_{1}f_{2}}{g_{1}e_{2}} + h.c.) \\
\end{split}    
\end{equation}
with transition dipole moment from the ground to the first-excited state $\mu_{e_{n}g_{n}}$ and from the first- to the second-excited state $\mu_{f_{n}e_{n}}$. These are usually assumed to be proportional $\mu_{f_{n}e_{n}} = \beta \mu_{e_{n}g_{n}}$, with scaling factor $\beta$. For simplicity, we assume that $\beta = 0$ in our model, so that higher-excited states are not populated by direct excitation. Furthermore, the transition dipole moments of the two chromophores are assumed to be parallel to each other.

In A-2DES, the system interacts with a sequence of four collinear laser pulses (Fig. \ref{fig:pulse_FD}a).
The electric field is given by:
\begin{equation}
    E(t) = \sum_{i=1}^{4} E_{i}(t)
\end{equation}
where each pulse is described as:
\begin{equation}
    {E}_{i}(t)
    = E_{i}^{0} \mathcal{E}_{i}(t-\tau_{i}) e^{i (\bm{k}_{i} \cdot \bm{r} - \omega_{i} t + \phi_{i})} +c.c.
\label{eq:field}
\end{equation}
characterized by amplitude $E_{i}^{0}$, envelope $\mathcal{E}_{i}(t)$, central time $\tau_{i}$, wavevector $\bm{k}_{i}$, frequency $\omega_{i}$ and phase $\phi_{i}$. The pulses are separated by the delay times $t_{1}$, $t_{2}$ and $t_{3}$ that are scanned during the experiment, while the emission of the incoherent signal is acquired during the detection time $t_{d}$ (Fig. \ref{fig:pulse_FD}a).

The A-2DES signal can be described at the fourth order in perturbation theory:\cite{pedromoortiz:2012, maly:2018, schroter:2018}
\begin{equation}
\begin{split}
    S^{(4)}(t) =& \int_{0}^{\infty} dt_{d} \int_{0}^{\infty} dt_{3} \int_{0}^{\infty} dt_{2} \int_{0}^{\infty} dt_{1} \, R^{(4)}(t_{d}, t_{3}, t_{2}, t_{1}) \\
    &\times E(t - t_{d}) E(t - t_{d} - t_{3}) E(t - t_{d} - t_{3} - t_{2}) \\
    &\times E(t - t_{d} - t_{3} - t_{2} - t_{1})
\end{split}
\label{eq:signal}
\end{equation}
where the fourth-order response function is:
\begin{equation}
\begin{split}
    &R^{(4)}(t_{d}, t_{3}, t_{2}, t_{1}) = \Tr\left\{ \hat{S} \rho^{(4)}(t_{d}, t_{3}, t_{2}, t_{1}) \right\} \\
    & =\left( \frac{i}{\hbar} \right)^{4} \Tr\left\{\hat{S} \super{G}(t_{d}) \super{M} \super{G}(t_{3}) \super{M} \super{G}(t_{2}) \super{M} \super{G}(t_{1}) \super{M} \rho(0) \right\}.
\end{split}
\label{eq:response}
\end{equation}
Initially, the system is in the common ground state $\rho(0) = \ketbra{G}{G}$. Each light-matter interaction is determined by the dipole moment superoperator $\super{M} \bigcdot = \left[\hat{\mu}, \bigcdot \right]$, while the dynamics during the delay times is given by the time-evolution superoperator $\super{G}(t) = \Theta(t) e^{\super{L} t}$, where $\Theta (t)$ is the Heaviside step function introduced to enforce causality.\cite{mukamel:1995}
Assuming that only exciton recombination contributes to the emission of the incoherent signal, the corresponding operator is given by the weighted sum of the excited-state populations:
\begin{equation}
\begin{split}
    \hat{S} =& k_{g_{1}e_{1}} \ketbra{e_{1}g_{2}}{e_{1}g_{2}} + k_{g_{2}e_{2}} \ketbra{g_{1}e_{2}}{g_{1}e_{2}} \\
    &+ (k_{g_{1}e_{1}} + k_{g_{2}e_{2}} ) \ketbra{e_{1}e_{2}}{e_{1}e_{2}}
\end{split}
\end{equation}
while the contribution of second-excited states is neglected as their relaxation is dominated by non-radiative internal conversion.
Furthermore, we assume the absence of any indirect mechanisms affecting the signal emission and detection.\cite{bolzonello:2023}

At $t = 0\ \si{fs}$, the signal is proportional to the excited-state population generated by the fourth pulse, which contain the same information of the third-order polarization of C-2DES.
However, in an A-2DES experiment, the spectroscopic observable is represented by the time-integrated signal:
\begin{equation}
    \overline{S} = \int_{0}^{+\infty} dt \ S(t).
\end{equation}

The different contributions to the optical response can be separated by controlling the pulse phases $\phi_{i}$, using either phase-cycling \cite{tian:2003, tan:2008} or phase-modulation \cite{tekavec:2007} schemes. In the following, we focus on the analysis of the rephasing signal, associated with the phase $\phi_{S} = - \phi_{1} + \phi_{2} + \phi_{3} - \phi_{4}$.
The delay times $t_{1}$ and $t_{3}$ are scanned from $0$ to $300\ \si{fs}$ in steps of $5\ \si{fs}$, while the waiting time $t_{2}$ is varied from $0$ to $100 \ \si{fs}$ in steps of $2 \ \si{fs}$. The signal is collected in a rotating frame at angular frequency $\omega_{\text{RF}} = 2.287\ \si{[rad]/fs}$ and then zero-padded.
By taking the Fourier transform along the delay times $t_{1}$ and $t_{3}$, a 2D spectrum is displayed as a function of $\hbar \omega_{1}$ and $\hbar \omega_{3}$ for each value of $t_{2}$.

Assuming time-ordering, impulsive limit, and the rotating-wave approximation, the optical response can be interpreted in terms of Feynman Diagrams (FDs),\cite{mukamel:1995} which represent the pathways followed by the system upon a specific sequence of light-matter interactions (Fig. \ref{fig:pulse_FD}b). These pathways can be distinguished into Ground-State Bleaching (GSB), Stimulated Emission (SE), and Excited-State Absorption (ESA). Depending on the final excited-state population, two types of ESA pathways can be identified, ending either in a one-exciton (ESAI) or in a two-exciton (ESAII) state.\cite{pedromoortiz:2012} 
The sign of each pathway is given by $(-1)^{m_{b}}$, where $m_{b}$ are the number of interactions on the \textit{bra} side of each FD. Therefore, GSB, SE, and ESAI contribute with negative signs, while ESAII contributes with a positive sign. 
According to this pathway decomposition, it is sometimes useful to express the A-2DES response as a simple weighted sum of the different components:
\begin{equation}
    R^{(4)} = R_{\text{GSB}} + R_{\text{SE}} + R_{\text{ESAI}} + \Gamma R_{\text{ESAII}}
    \label{eq:componenti}
\end{equation}
where $\Gamma$ is a weight determined by the dynamics of the excited-state populations during the detection time $t_{d}$.\cite{pedromoortiz:2012, maly:2018, schroter:2018, kunsel:2019, bruschi:2023} For our model, $\Gamma=2$ when the only decay channel is exciton recombination, and in this particular case we recover the C-2DES spectra. In all the other cases, the discrepancies between coherent- and action-detected spectra result from the different weights of the pathways in the two techniques.
The detailed expressions for each class of pathways are reported in the Supplementary Material.
Note that pathways of opposite signs can interfere destructively in the spectrum, leading to partial or complete cancellation of the associated spectral features.

As shown schematically in Fig. \ref{fig:pulse_FD}c, spectral features of a molecular dimer may appear both as diagonal peaks (D1 and D2) and cross peaks (C1 and C2). In order to interpret the results discussed in the next section, we note that, in the considered model, ESA-type pathways contribute to cross-peak positions, while GSB and SE pathways appear both as diagonal and cross peaks. For completeness, the spectra and amplitudes associated with each class of pathways are reported in the Supplementary Material.  

\section{Results and Discussion}

\subsection{Spectral features at early-waiting times}
\label{sec:IIIA}
\begin{figure*}
    \centering
    \includegraphics[width=\textwidth]{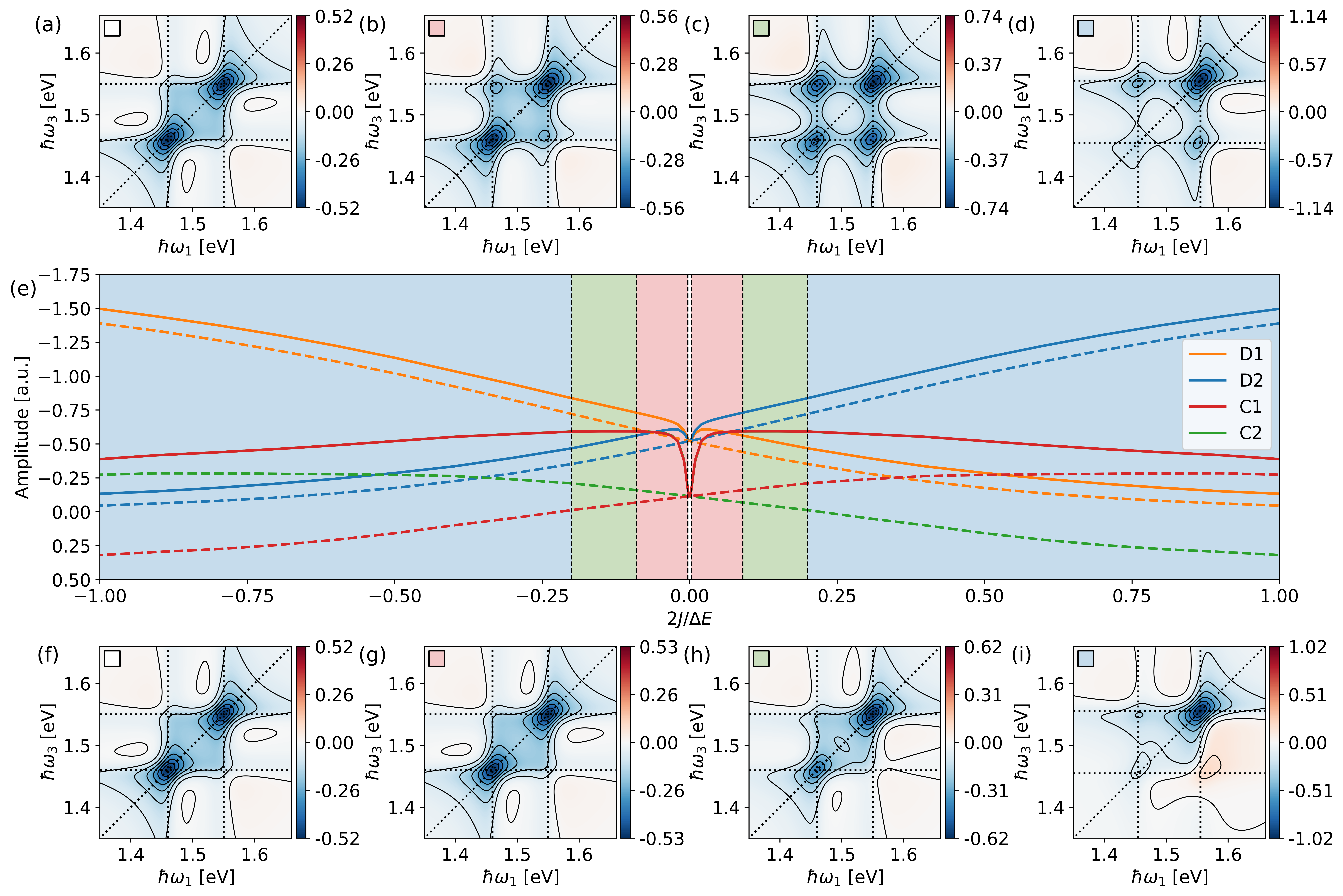}
    \caption{(a-d) A-2DES spectra for different values of $2J/\Delta E$: (a) $0.0$, (b) $0.005$, (c) $0.1$, (d) $0.5$. (e) Comparison between the amplitudes of the spectral features as a function of $2J / \Delta E$ in A-2DES (solid lines) and C-2DES (dashed lines). The different regions correspond to: (white region) no cross peaks, (red region) incoherent mixing, (green region) ESA-free, (blue region) delocalization. (f-i) C-2DES spectra for different values of $2J/\Delta E$: (f) $0.0$, (g) $0.005$, (h) $0.1$, (i) $0.5$.}
    \label{fig:profile}
\end{figure*}

We start by analyzing the spectral features that appear in the rephasing signal of A-2DES at waiting times $t_{2} = 0 \ \si{fs}$ for different values of the excitonic coupling $J$ (Fig. \ref{fig:profile}a-d).
In Fig. \ref{fig:profile}e, the amplitudes of the diagonal and cross peaks corresponding to A-2DES (solid lines) are shown as a function of $2J/\Delta E$, where $\Delta E = \epsilon_{e_{2}} - \epsilon_{e_{1}}$ is the energy difference between the site energies of the two chromophores. 
The amplitude of the spectral features is extracted from the spectra at frequencies corresponding to that of one-exciton eigenstates: $\hbar \omega_{E_{k}G} = \epsilon_{E_{k}} - \epsilon_{G}$. 

The behavior of diagonal and cross peaks as a function of the excitonic coupling prompts the identification of different regions (Fig. \ref{fig:profile}e), each associated with distinct key mechanisms determining the amplitude of the spectral features.
In the absence of excitonic coupling, the system is composed of two independent chromophores and the spectrum only exhibits diagonal peaks while cross peaks are not present (Fig. \ref{fig:profile}a). 
Since the transition dipole moment from the first to higher-excited states has been set to zero, diagonal peaks are only given by GSB and SE contributions, which represent the non-linear optical response of individual chromophores. 
The absence of cross peaks results from the complete cancellation between positive features of ESAII and negative features of GSB, SE, and ESAI pathways at these positions, which happens when $\Gamma=2$ in Eq. \eqref{eq:componenti}. 
This regime is associated with the cusp in the amplitude of spectral features at $2J/\Delta E = 0$ (white region in Fig. \ref{fig:profile}e).

As soon as the excitonic coupling becomes non-zero, the EEA channel is activated with a specific rate which depends on the excitonic coupling and the rate of the internal conversion processes. This additional relaxation process, occurring during the detection time, is responsible for the appearance of cross peaks in the spectrum (Fig. \ref{fig:profile}b).
In general, this can be understood as an incomplete cancellation of the negative contributions from GSB, SE, and ESAI by the positive ESAII features, whose amplitude is reduced by the EEA process.\cite{maly:2018, schroter:2018, kunsel:2019, bruschi:2023}
Alternatively, it can be attributed to the incoherent mixing of linear signals of individual chromophores caused by non-linear population dynamics, i.e., EEA, during the detection time.\cite{gregoire:2017, kalaee:2019, bruschi:2023} Depending on the rate of the EEA compared to signal emission, the amplitude of cross peaks ranges from zero to a maximum value (red region in Fig. \ref{fig:profile}e).
The amplitude of diagonal peaks also exhibits a slight increase due to constructive interference with the emerging cross peaks in the spectrum.
In this region, cross peaks do not provide information about the non-linear response of the system, but rather reflect incoherent mixing during the detection time. However, it represents an important signature of interaction between the two chromophores.
Note that this regime was not considered in previous studies of A-2DES, as EEA was introduced as an incoherent process without explicitly accounting for the delocalization between states in the two-exciton manifold \cite{maly:2018, schroter:2018}.

When the amplitude of the cross peaks is at its maximum value (Fig. \ref{fig:profile}c), it remains approximately constant in a certain range of excitonic couplings (green region in Fig. \ref{fig:profile}e). Indeed, when annihilation is complete, the spectrum result ESA-free and the amplitude of cross peaks is determined only by the GSB and SE contributions, namely by the product of the transition dipole moments $\abs{\mu_{E_{1}G}}^{2} \abs{\mu_{E_{2}G}}^{2}$. When the transition dipole moments of the two chromophores are equal, the product is independent of $J$ to first order, as explicitly reported in the Supplementary Material.  
On the other hand, the relative amplitude of the diagonal peaks already reflects the redistribution of the transition dipole moment, as they are proportional to $\abs{\mu_{E_{n}G}}^{4}$ . Indeed, excitonic delocalization within the one-exciton manifold causes one eigenstate to become progressively brighter while the other becomes darker. For positive excitonic coupling ($J > 0$), the upper diagonal peak increases in amplitude while the lower diagonal peak decreases. The opposite trend is observed for negative excitonic coupling ($J < 0$).
In addition, the increase in excitonic coupling leads to a progressive shift in the peak position, reflecting excitonic splitting between states in the one-exciton manifold.
For larger excitonic coupling, cross peaks are also affected by the transition dipole moment redistribution (Fig. \ref{fig:profile}d), showing a decrease in amplitude (blue region in Fig. \ref{fig:profile}e).

It is interesting to compare these results with the one obtained in C-2DES (Fig. \ref{fig:profile}f-i) for the same model system. Although the amplitudes of the diagonal peaks are comparable, significant differences in the behavior of the cross-peak amplitudes are evident especially in the region of weak to intermediate excitonic coupling, as already reported in previous studies.\cite{maly:2018, schroter:2018} Notably, the amplitude of cross peaks in A-2DES reaches its maximum already at weak coupling, while in C-2DES grows linearly with the coupling strength because of the different interplay of ESA-type pathways with GSB and SE in the two signals.

Regarding the significance of cross peaks in A-2DES, we want to highlight that they convey different information depending on the excitonic coupling strength. In the weak-coupling regime (red region in Fig. \ref{fig:profile}e), cross peaks merely represent a signature of EEA and can be regarded as resulting from the incoherent mixing of the linear spectra of individual chromophores.\cite{bruschi:2023} At intermediate excitonic coupling (green region in Fig. \ref{fig:profile}e), the amplitude of cross peaks is maximum, since the spectrum is ESA-free. However, the nature of the states being probed differs from the non-interacting limit. In these conditions, the cross peaks evolution can reveal excited-state dynamics with high contrast, as discussed in the next Section. In the regime of strong excitonic coupling, the transition dipole moment redistribution accompanying exciton delocalization starts also to affect cross peaks, whose amplitude decreases until it reaches that of the cross peak observed in C-2DES.

\subsection{Waiting-time dynamics and coherent oscillations}
\label{sec:IIIB}

\begin{figure*}
    \centering
    \includegraphics[width=\textwidth]{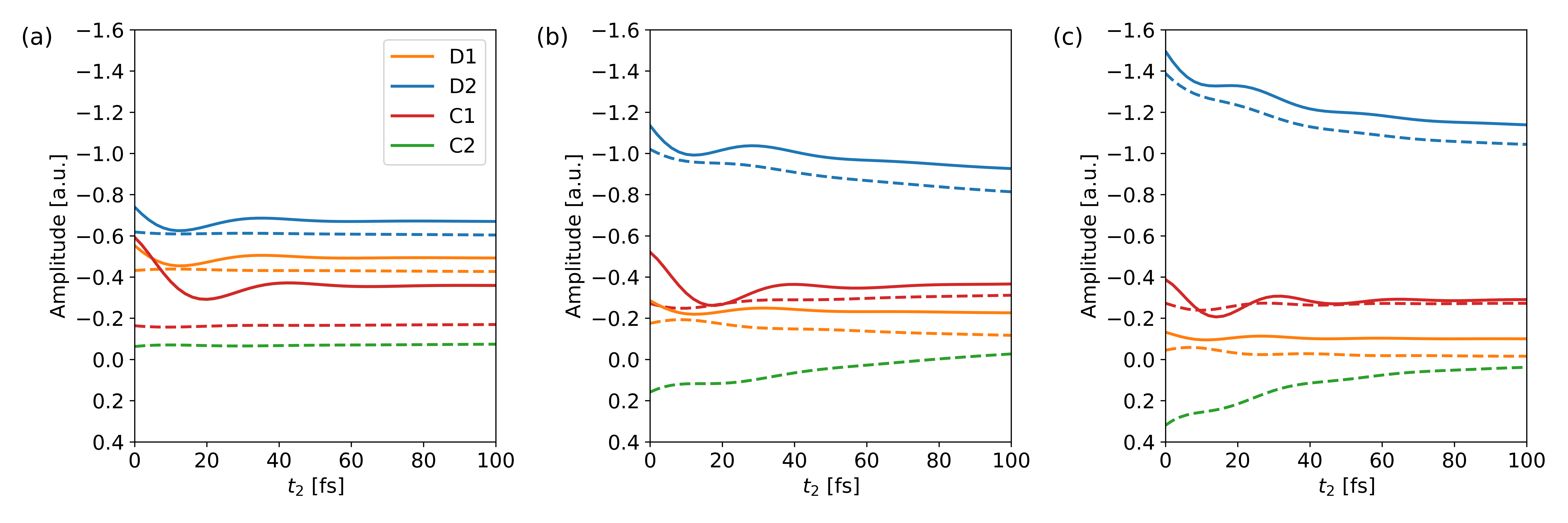}
    \caption{Amplitudes of spectral features in A-2DES (solid lines) and C-2DES (dashed lines) as a function of the waiting time $t_{2}$ for (a) $2J/\Delta E = 0.1$, (b) $2J/\Delta E = 0.5$, and (c) $2J/\Delta E = 1.0$.}
    \label{fig:t2_dynamics}
\end{figure*}

In the following, we investigate the dynamics of spectral features in A-2DES during the waiting time $t_{2}$.
Specifically, we focus on the short-time coherent evolution of the system, without considering long-time equilibration processes.

In Fig. \ref{fig:t2_dynamics}, the amplitude of spectral features is shown as a function of $t_{2}$ for different values of excitonic coupling for both A-2DES and C-2DES.
In A-2DES (solid lines), we observe coherent oscillations in the amplitudes of both diagonal and cross peaks for the considered values of excitonic coupling.
The presence of these oscillations is associated with the preparation of a coherent superposition of eigenstates upon ultrafast excitation.\cite{mancal:2020}
In our model, the rephasing pathways (Fig. \ref{fig:pulse_FD}b) contributing to cross-peak oscillate during the waiting time, reflecting the evolution of the coherence between one-exciton eigenstates.
Conversely, the oscillations in the diagonal peak originate both from the overlap with oscillating cross peaks and from non-secular contributions to the dynamics, which couple the evolution of populations and coherences in the eigenstate basis.

As the excitonic coupling increases, we observe a more pronounced oscillating behavior in the amplitude of spectral features.
On the one hand, the oscillation frequency $\omega_{E_{2} E_{1}} = (\epsilon_{E_{2}} - \epsilon_{E_{1}})/\hbar$ increases due to the larger excitonic splitting between eigenstates. 
On the other hand, the pure-dephasing rate between one-exciton eigenstates decreases as a result of excitonic delocalization.\cite{mancal:2020}
Therefore, these effects lead to faster and longer-living coherent oscillations as the excitonic coupling increases. This behavior is confirmed by the power spectra obtained from the Fourier transform of the amplitudes over the waiting time $t_{2}$, as shown in the Supplementary Material.

In Fig. \ref{fig:t2_dynamics}, we also report the amplitude of the spectral features of C-2DES (dashed lines). We notice a striking difference in the amplitude of the oscillations compared to the previous case, especially when the excitonic coupling is weaker (Fig. \ref{fig:t2_dynamics}a).
To explain the difference between the two techniques, we examine the pathways that are in a coherence during $t_{2}$.
Assuming an ESA-free spectrum (green and blue regions in Fig. \ref{fig:profile}), coherent oscillations are only due to SE pathways in A-2DES. The amplitude of these pathways is:
\begin{equation}
\begin{split}    
    R_{\text{SE}} &\propto \mu_{E_{n}G}^2 \mu_{E_{m}G}^2 \\ 
    &= \mu_{e_{1}g_{1}}^2 \mu_{e_{2}g_{2}}^2 + O(J)
\end{split}
\end{equation}
resulting in a lowest-order contribution that is independent of excitonic coupling.
In contrast, the appearance of coherent oscillations in C-2DES is determined by interference between SE and ESA pathways, which have opposite signs. In this case, the amplitude of the resulting contribution is: 
\begin{equation}
\begin{split}
    R_{\text{SE}} + R_{\text{ESA}}  \propto & \mu_{E_{n}G}^2 \mu_{E_{m}G}^2 - \mu_{E_{n}G} \mu_{E_{m}G} \mu_{F_{l}E_{n}} \mu_{F_{l}E_{m}} \\
    = & J (\mu_{e_{1}g_{1}}^3 \mu_{e_{2}g_{2}} - \mu_{e_{1}g_{1}} \mu_{e_{2}g_{2}}^3) + \frac{2 J^2}{\Delta E^2} \\
    &\times \left[\mu_{e_{1}g_{1}}^4 + \mu_{e_{2}g_{2}}^4 - (2 - \alpha ^2)\mu_{e_{1}g_{1}}^2 \mu_{e_{2}g_{2}}^2\right] + O\left(J^{3}\right)
\end{split}
\end{equation}
which is linear at the lowest order in the excitonic coupling if the transition dipole moments of the chromophores are different and quadratic if they are equal.
Therefore, in C-2DES, coherent oscillations emerge only as a result of excitonic delocalization, while in A-2DES they can appear as soon as EEA becomes effective.
As a consequence, A-2DES can display coherent oscillations even at weak excitonic coupling, where they are absent in C-2DES due to the reduced amplitude of cross peaks.

\begin{figure*}
    \centering
    \includegraphics[width=0.9\linewidth]{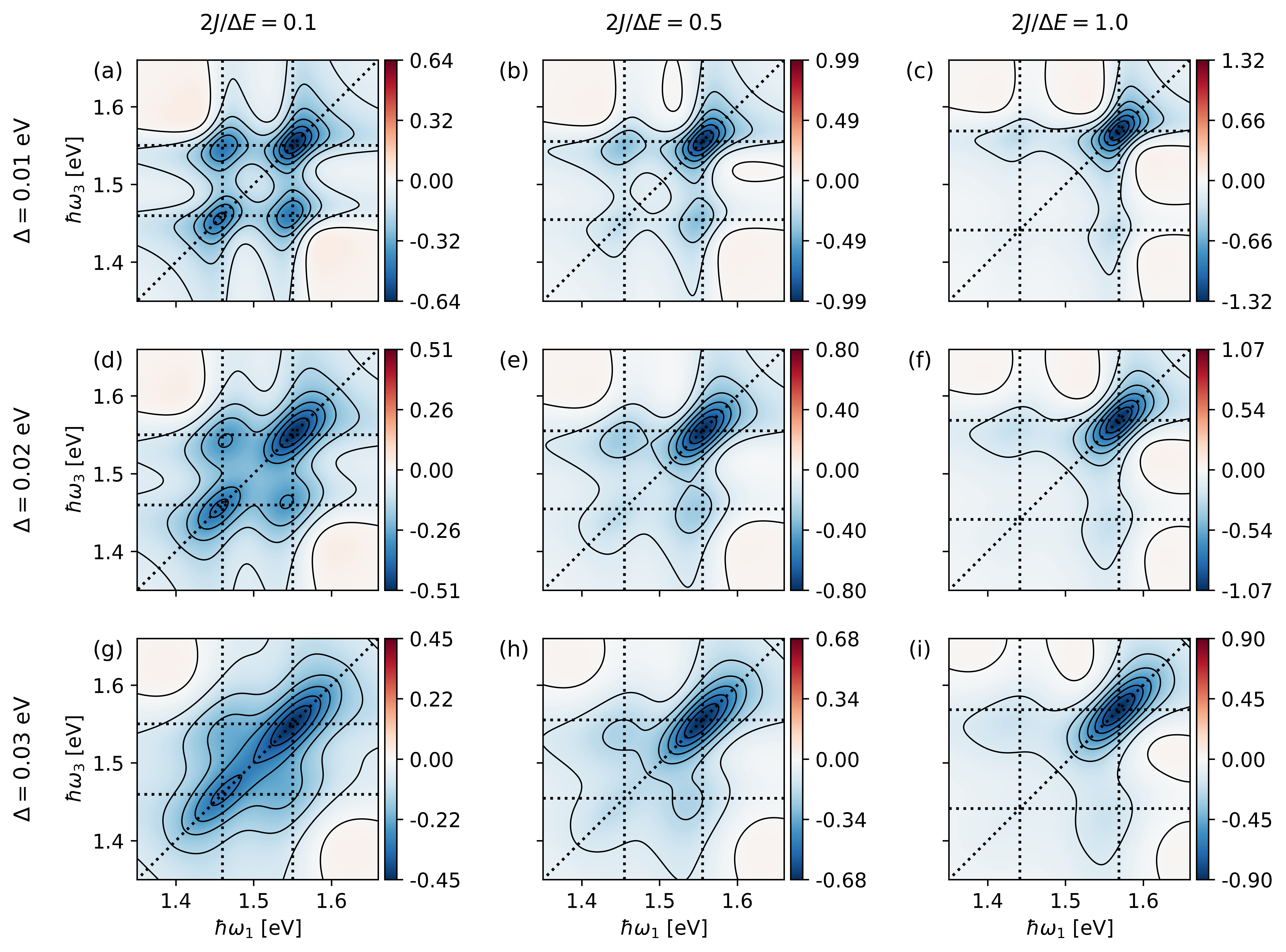}
    \caption{A-2DES spectra for different values of excitonic coupling (columns) and static disorder (rows).}
    \label{fig:disorder_spectra}
\end{figure*}

Coherent oscillations in cross-peak positions have been extensively studied in C-2DES and may arise from a variety of physical mechanisms, depending on the system under investigation. For instance, these can arise because of electronic and vibronic coherences,\cite{butkus:2012}, ground and excited-state wavepacket evolution,\cite{tiwari:2013} and coherences between polaritonic states.\cite{peruffo:2023, toffoletti:2025}

In our purely-electronic system, the observed oscillations arise from the evolution of a coherence between one-exciton eigenstates.
Depending on the strength of excitonic coupling, these eigenstates may correspond either to (localized) site states, where oscillations represent inter-site coherences, or (delocalized) excitonic states, where oscillations indicate inter-excitonic coherences.\cite{kiessling:2020}
However, it should be noticed that the larger amplitude of coherent oscillation in A-2DES compared to C-2DES is intrinsically tied to the different mechanism of pathway cancellation between the two techniques. Therefore, we argue that this enhancement is independent of the physical origin of the coherent dynamics. In small multi-chromophoric systems, where the contribution of incoherent mixing does not completely dominate the spectrum, the higher cross-peak amplitude in A-2DES offers the opportunity to measure coherent oscillations with improved contrast compared to C-2DES.

However, the presence of cross peaks is a necessary but not sufficient condition for observing coherent oscillations during the waiting time.
Indeed, the ability to experimentally resolve such oscillations also depends on the finite bandwidth of pulses, as their duration must be shorter than the ultrafast dynamics being probed.
Moreover, when pulses overlap significantly, different time orderings of the light-matter interaction are possible, giving rise to additional pathways.\cite{rose:2021} 
This phenomenon, known as coherent artifact,\cite{palecek:2019} can obscure the coherent dynamics under consideration.
To investigate the effect of finite pulse bandwidth on the waiting-time dynamics, we also employed a non-perturbative approach to the light-matter interaction to simulate the optical response.\cite{bruschi:2022} As shown in the Supplementary Material, we verified that coherent oscillations become less pronounced and eventually disappear as the pulse duration increases.

In the considered model, coherent oscillations are consistently more pronounced in A-2DES compared to C-2DES for the same value of excitonic coupling.
The difference between the amplitude of coherent oscillations in the two techniques was also observed experimentally in squaraine dimers.\cite{maly:2020}
However, this has been attributed to technical aspects, namely the inherent phase stability of collinear A-2DES setup compared to the non-collinear C-2DES setup.
Instead, we emphasize that the different sensitivity to coherent dynamics of the two techniques is already encoded at the level of the response function.
Nevertheless, since pathways involving an inter-site coherence rapidly dephase, their contribution usually manifested as a fast decay in the amplitude of the signal along the waiting time, as recently reported for the LH2 complex.\cite{javed:2024}
Together with previous results,\cite{maly:2020, javed:2024} our findings suggest a higher sensitivity of A-2DES in detecting coherent superposition of eigenstates compared to C-2DES.

\subsection{Static Disorder and Broadening of the Spectral Features}
\label{sec:IIIC}

In the following, we introduce static disorder in the system to investigate the effect of spectral broadening in A-2DES.
Specifically, we consider only diagonal disorder, such that the site energy of each chromophore is given by $\epsilon_{e_{n}}^{'} = \epsilon_{e_{n}} + \delta \epsilon_{e_{n}}$, where the static energy fluctuation $\delta \epsilon_{n}$ is randomly sampled from a Gaussian distribution with zero mean and standard deviation $\Delta$.
For each realization of static disorder, a spectrum is simulated. Then, the overall spectrum of the inhomogeneous sample is obtained by averaging over $10^{4}$ different realizations.

We focus on cases where EEA is complete during the detection time, i.e., ESA-free spectra (green and blue regions in Fig. \ref{fig:profile}e). 
In Fig. \ref{fig:disorder_spectra}, we report the A-2DES spectra corresponding to different values of excitonic coupling and static disorder.
As static disorder increases, we observe changes in the lineshapes of spectral features: diagonal peaks become elongated along the diagonal, while cross peaks appear more rounded due to inhomogeneous broadening.
For a fixed level of disorder, the effect of excitonic coupling on the lineshape is less pronounced and more difficult to evaluate qualitatively because of the difference in amplitude between diagonal and cross peaks.

\begin{figure}
    \centering
    \includegraphics[width=\columnwidth]{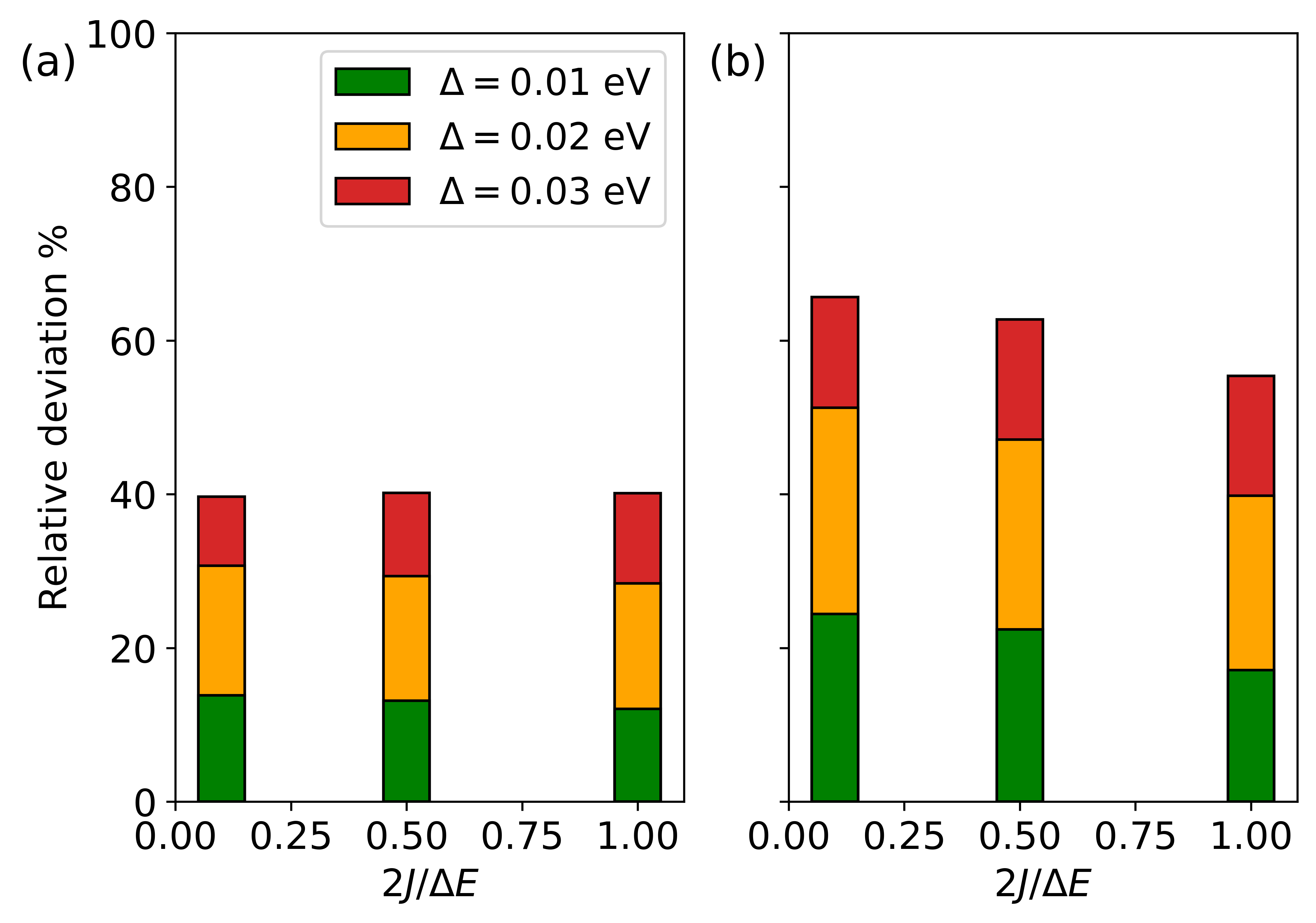}
    \caption{Relative deviation of (a) diagonal-peak and (b) cross-peak amplitude between inhomogeneous and homogeneous cases for different values of static disorder $\Delta$ and excitonic coupling $2J/\Delta E$ = 0.1, 0.5, 1.0.}
    \label{fig:disorder}
\end{figure}

To quantify and visualize these changes, we consider the deviation in the relative amplitude of the diagonal peak (Fig. \ref{fig:disorder}a) and cross peaks (Fig. \ref{fig:disorder}b) between the inhomogeneous and homogeneous cases for different excitonic couplings. 
The relative deviation is defined as $\abs{A_{\text{inh}} - A_{\text{hom}}}/\abs{A_{\text{hom}}} \cdot 100\%$, where $A_{\text{inh}}$ and $A_{\text{hom}}$ are the peak amplitudes in the case of inhomogeneous and homogeneous broadening, respectively.
As static disorder increases, both diagonal and cross peaks exhibit larger relative deviations, indicating an overall reduction in the amplitude.
However, cross peaks are more influenced by static disorder compared to the diagonal peak, as evidenced by the higher relative deviation for the same value of static disorder.

Moreover, we observe a different behavior of diagonal and cross peaks features with respect to their dependence on the excitonic coupling.
As the excitonic coupling increases, the relative deviation remains approximately constant for diagonal peaks, while it decreases for cross peaks.
This suggests that cross peaks are more robust to static inhomogeneities in the case of strong coupling compared to weak coupling.  

These observations can be rationalized within the framework of stochastic lineshape theory.
We assume that transition frequency fluctuations are uncorrelated between different chromophores, while those of each chromophore are characterized by a time-correlation function $C(t) = \gamma \delta(t) + 2 \sigma^{2}$, consisting of white-noise and static components. 
The fluctuation amplitudes are defined as $\gamma = k_{\text{D}}/2 \hbar$ and $\sigma = \Delta/\sqrt{2}\hbar$, based on the parameters of the model (see Tab. \ref{tab:tableI}).
This description leads to a Voigt lineshape function $g(t) = \gamma t + \sigma^{2} t^{2}$ for the transition, where the two terms reflect homogeneous and inhomogeneous broadening, respectively.\cite{hamm:2011}
For convenience, we define the mixing angle $\theta = \frac{1}{2} \tan^{-1} \left(\frac{2J}{\Delta E} \right)$, such that $0 \leq \theta \leq \pi/4$.
With these definitions and by considering only the pure-dephasing contribution to the lineshape, the dephasing function of pathways contributing to diagonal peaks can be written as:
\begin{equation}
\begin{split}
    F_{\text{diag}}(t_{3}, t_{2}, t_{1}) =& e^{-\gamma \left(1 - 2 \sin^{2}\theta \cos^{2}\theta \right) (t_{1} + t_{3})} \\
    & \times e^{- \sigma^{2} \left( 1 - 2 \sin^{2}\theta \cos^{2}\theta \right) (t_{1} - t_{3})^2}.
    \label{eq:F_diag}
\end{split}
\end{equation}
Note that for $t_{1} = t_{3}$, these pathways will always rephase, resulting in peaks elongated along the diagonal and characterized by homogeneous broadening in the anti-diagonal direction.
On the other hand, the dephasing function for pathways contributing to cross peaks is:
\begin{equation}
\begin{split}
    F_{\text{cross}}(t_{3}, t_{2}, t_{1}) =& e^{-\gamma \left(1 - 2 \sin^{2}\theta \cos^{2}\theta \right) (t_{1} + t_{3})} \\
    & \times e^{- \sigma^{2} \left[ \left( t_{1}^2 + t_{3}^2 \right) - 2 \sin^{2}\theta \cos^{2}\theta (t_{1} + t_{3})^2 \right]}
    \label{eq:F_cross}
\end{split}
\end{equation}
which exhibit a more complex dependence on the mixing angle. In this regard, it is instructive to consider two limiting cases.
In the weak-coupling limit ($\theta \rightarrow 0$), the lineshape function related to cross peaks is equivalent to that of the product of the linear spectra of two different chromophores.
In this regime, cross peaks lack rephasing capability and thus appear rounded in the presence of inhomogeneous broadening.\cite{yang:1999, bolzonello:2023}
Conversely, in the strong-coupling limit ($\theta \rightarrow \pi/4$), cross peaks also rephase for $t_{1} = t_{3}$, leading to diagonally elongated features similarly to diagonal peaks.
Between these two limits, cross peaks will have only partial rephasing capability.

Therefore, we arrive at the somehow unexpected conclusion that inhomogeneous broadening plays a positive role in reducing the amplitude of cross peaks related to incoherent mixing.
In addition, analyzing the lineshape may help in discerning whether cross peaks only reflect incoherent mixing or are indicative of excitonic delocalization.

\section{Conclusions}

In this work, we investigated the effect of excitonic coupling on the spectral features of A-2DES, focusing on a molecular dimer model.
By explicitly including higher-excited states of each chromophore, we accounted for exciton-exciton annihilation arising from delocalization within the two-exciton manifold.
Our results reveal changes in the A-2DES spectra ranging from the non-interacting case, characterized by the absence of cross peaks, to the strong-coupling regime, where cross peaks provide information about excitonic delocalization similar to C-2DES, but with a distinct amplitude pattern.
In the weak to intermediate coupling regime, the significance of cross peaks depends on the strength of the excitonic coupling, leading us to propose a classification based on the cross-peak amplitude. 
Notably, A-2DES exhibits significant cross peaks even under weak coupling, providing higher contrast for observing coherent oscillations during the waiting time compared to C-2DES.
Additionally, we considered the effect of inhomogeneous broadening, demonstrating that cross peaks arising from incoherent mixing are more susceptible to static disorder, whereas those associated with excitonic delocalization are more robust.
Based on these findings, we suggest that A-2DES is a powerful technique for probing excited-state dynamics in small multi-chromophoric systems, especially in the intermediate-coupling regime. 
However, we remark that in larger multi-chromophoric systems, the spectrum can be dominated by the contribution of incoherent mixing which may obscure excited-state dynamics.\cite{bolzonello:2023, lopez_ortiz:2024, javed:2024}
Recently, several methods have been proposed to remove the contribution of incoherent mixing, including time gating\cite{maly:2018, kunsel:2019, bruschi:2022} and frequency gating\cite{yang:2023} of the signal emission, polarization-controlled laser pulses, \cite{faitz:2024} and using spectro-temporal symmetries in the A-2DES signal.\cite{charvatova:2025}.
In this regard, static disorder and excitonic delocalization may still help in mitigating the unfavorable balance of the non-linear response compared to incoherent mixing.\cite{javed:2024}

\section*{Supplementary Material}
Analytical characterization of the molecular dimer model, additional figures and analysis of the individual components of the fourth-order response function, Fourier power spectra of the waiting-time dynamics, and results of non-perturbative simulations. 

\begin{acknowledgments}
The authors want to thank Elisabetta Collini for the fruitful discussion about the work.
M.B, F.G. and B.F. acknowledge the financial support by the Department of Chemical Sciences (DiSC) and the University of Padova with Project QA-CHEM (P-DiSC No. 04BIRD2021-UNIPD).
R.Z. acknowledges MIUR “Dipartimenti di Eccellenza” under the project Chemical Complexity (C$^{\text{2}}$) for funding his Ph.D. scholarship.
B.F. acknowledges funding from the European Union - NextGenerationEU, within the National Center for HPC, Big Data and Quantum Computing (Project no. CN00000013, CN1 Spoke 10: “Quantum Computing”).
The authors acknowledge C3P (Computational Chemistry Community in Padua) HPC facility of the Department of Chemical Sciences (DiSC) of the University of Padova.
\end{acknowledgments}

\section*{Author Declarations}

\subsection*{Conflict of Interest}
The authors have no conflicts to disclose.

\subsection*{Author contributions}
M.B. and R.Z. contributed equally to this work.

\section*{Data Availability}

The data that support the findings of this study are available from the corresponding authors upon reasonable request.

\bibliography{bibliography}

\end{document}


\maketitle

\section{Eigenstates and eigenvalues}

For convenience, we define the quantities in Table \ref{tab:diagonalization}. By diagonalizing the Hamiltonian relative to the one-exciton manifold, we obtain the following eigenstates:
\begin{equation}
\begin{split}
    \ket{E_{1}} &= \cos{\theta}\ket{e_{1}g_{2}} - \sin{\theta}\ket{g_{1}e_{2}} \\
    \ket{E_{2}} &= \sin{\theta}\ket{e_{1}g_{2}} + \cos{\theta}\ket{g_{1}e_{2}}
\label{eq:one exciton states}
\end{split}
\end{equation}
and the eigenvalues:
\begin{equation}
\begin{split}
    E_{E_{1}} &= \Bar{E} - \Omega_{0} \\
    E_{E_{2}} &= \Bar{E} + \Omega_{0}
\end{split}
\end{equation}
Instead, by diagonalizing the Hamiltonian of the two-exciton manifold, we obtain the following eigenstates:
\begin{equation}
\begin{split}
    \ket{F_{1}} &= - \frac{K}{\Omega}\ket{e_{1}e_{2}} + \frac{\Delta E + \Omega}{2\Omega}\ket{f_{1}g_{2}} - \frac{\Delta E - \Omega}{2\Omega}\ket{g_{1}f_{2}} \\
    \ket{F_{2}} &= \frac{\Delta E}{\Omega}\ket{e_{1}e_{2}} + \frac{K}{\Omega}\ket{f_{1}g_{2}} - \frac{K}{\Omega}\ket{g_{1}f_{2}} \\
    \ket{F_{3}} &= \frac{K}{\Omega}\ket{e_{1}e_{2}} - \frac{\Delta E - \Omega}{2\Omega}\ket{f_{1}g_{2}} + \frac{\Delta E + \Omega}{2\Omega}\ket{g_{1}f_{2}}
\end{split}
\label{eq:two exciton states}
\end{equation}
and the corresponding eigenvalues:
\begin{equation}
\begin{split}
    E_{F_{1}} &= 2\Bar{E} - \Omega \\
    E_{F_{2}} &= 2\Bar{E} \\
    E_{F_{3}} &= 2\Bar{E} + \Omega
\end{split}
\end{equation}
\begin{table}[h]
\centering
\caption{\label{tab:diagonalization}Parameters used for defining eigenstates and eigenvalues.}
\begin{tabular}{c c c c}
\hline
\hline
Parameter & Definition & Parameter & Definition \\
\hline
$\Bar{E}$ & $\frac{\epsilon_{e_{1}} + \epsilon_{e_{2}}}{2}$ & $\Delta E$ & $\epsilon_{e_{2}} - \epsilon_{e_{1}}$ \\
$\Omega_{0}$ & $\sqrt{\left (\frac{\Delta E}{2} \right)^{2} + J^{2}}$ & $\Omega$ & $\sqrt{\Delta E^{2} + 2K^{2}}$ \\
$\cos{2\theta}$ & $\frac{\Delta E}{2 \Omega_{0}}$ & $\sin{2\theta}$ & $\frac{J}{\Omega_{0}}$\\
\hline
\hline
\end{tabular}
\end{table}

\section{Transition dipole moment in the eigenstate basis}

The expressions for transition dipole moment in the excitonic basis can be obtained upon change of basis  using the matrix of the eigenvectors reported in Eqs. \ref{eq:one exciton states} and \ref{eq:two exciton states}.
For the transition from the ground state to the one-exciton manifold are:
\begin{equation}
\mu_{E_{1}G} = \mu_{e_{1}g_{1}} \cos{\theta} - \mu_{e_{2}g_{2}} \sin{\theta} \qquad \qquad \qquad
\mu_{E_{2}G} = \mu_{e_{1}g_{1}} \sin{\theta} + \mu_{e_{2}g_{2}} \cos{\theta} 
\label{eq:dipole moments}
\end{equation}
while for the transition from the one-exciton to the two-exciton manifold are:
\begin{equation}
\begin{aligned}
\mu_{F_{1}E_{1}} &= -\dfrac{K}{\Omega} (- \mu_{e_{1}g_{1}} \sin{\theta} + \mu_{e_{2}g_{2}} \cos{\theta})
\hspace{2cm}
&\mu_{F_{1}E_{2}} &= -\dfrac{K}{\Omega} (\mu_{e_{1}g_{1}} \cos{\theta} + \mu_{e_{2}g_{2}} \sin{\theta}) \\
\mu_{F_{2}E_{1}} &= \dfrac{\Delta E}{\Omega} (- \mu_{e_{1}g_{1}} \sin{\theta} + \mu_{e_{2}g_{2}} \cos{\theta})
\hspace{2cm}
&\mu_{F_{2}E_{2}} &= \dfrac{\Delta E}{\Omega} (\mu_{e_{1}g_{1}} \cos{\theta} + \mu_{e_{2}g_{2}} \sin{\theta}) \\
\mu_{F_{3}E_{1}} &= \dfrac{K}{\Omega} (- \mu_{e_{1}g_{1}} \sin{\theta} + \mu_{e_{2}g_{2}} \cos{\theta})
\hspace{2cm}
&\mu_{F_{3}E_{2}} &= \dfrac{K}{\Omega} (\mu_{e_{1}g_{1}} \cos{\theta} + \mu_{e_{2}g_{2}} \sin{\theta})
\end{aligned}
\label{eq:dipole_moments_1}
\end{equation}

\section{Relaxation rates from the two-exciton to the one-exciton manifold}

By substituting the coefficients of the eigenstates in Eqs. \ref{eq:one exciton states} and \ref{eq:two exciton states} inside Eq. 11 of the main text, we derived the analytical expressions for the relaxation rates from two- to one-exciton manifold:

\begin{equation}
\begin{split}
    k_{E_{1} F_{1}} &= k_{E_{2} F_{3}} = k_{ER} \frac{K^2}{\Omega^2} + \frac{k_{IC}}{4} \left(\frac{\Delta E^2}{\Omega^2} + 1 + \frac{\Delta E^2}{\Omega_{0}\Omega }\right) \\
    k_{E_{1} F_{2}} &= k_{E_{2} F_{2}} = k_{ER} \frac{\Delta E^2}{\Omega^2} + k_{IC} \frac{K^2}{\Omega^2}  \\
    k_{E_{1} F_{3}} &= k_{E_{2} F_{1}} = k_{ER} \frac{K^2}{\Omega^2} + \frac{k_{IC}}{4} \left(\frac{\Delta E^2}{\Omega^2} + 1 - \frac{\Delta E^2}{\Omega_{0}\Omega}\right).
\end{split}
\label{eq:EEA rates}
\end{equation}
Considering the point of view of $\ket{E_{1}}$, in the absence of delocalization we have $k_{E_{1} F_{1}} = k_{IC}$, $k_{E_{1} F_{2}} = k_{ER}$ and $k_{E_{1} F_{3}} = 0$. However, in the strong-coupling regime $\ket{F_{2}}$ inherits the rapid IC character of $\ket{f_{n} g_{m}}$ and $k_{E_{1} F_{2}} = k_{IC}/2$. Consequently, $k_{E_{1} F_{1}}$ becomes equal to $k_{E_{1} F_{3}} = k_{ER}/2 + k_{IC}/4$.

\section{Inverse Partecipation Ratio}

The amount of excitonic delocalization is quantified by the Inverse Participation Ratio $\text{IPR}_{k} = \sum_{n} \abs{c_{nk}}^{4}$. In particular, its reciprocal $\text{IPR}_{k}^{-1}$ gives an estimate of the number of states over which the eigenstate is delocalized. If an eigenstate is completely localized on a single state, the $\text{IPR}^{-1}$ takes its minimum value. In the case of an eigenstate which is uniformly delocalized across the manifold, the $\text{IPR}^{-1}$ takes its maximum value equal to the number of site states in the manifold.
Using the eigenstates in Equations \ref{eq:one exciton states} and \ref{eq:two exciton states}, the IPR can be written as follows:
\begin{equation}
\begin{split}
    \text{IPR}_{E_{1}} &= \text{IPR}_{E_{2}} = 1 - \frac{J^2}{2\Omega_{0}^2} \\
    \text{IPR}_{F_{1}} &= \text{IPR}_{F_{3}} = \frac{2 \Delta E^4 + 3 K^4 + 4 \Delta E^2K^2}{2 \Omega^4} \\
    \text{IPR}_{F_{2}} &= \frac{\Delta E^4 + 2 K^4}{\Omega^4}
\end{split}
\label{eq:IPR}
\end{equation}

\begin{figure}[H]
    \centering
    \includegraphics[width=\linewidth]{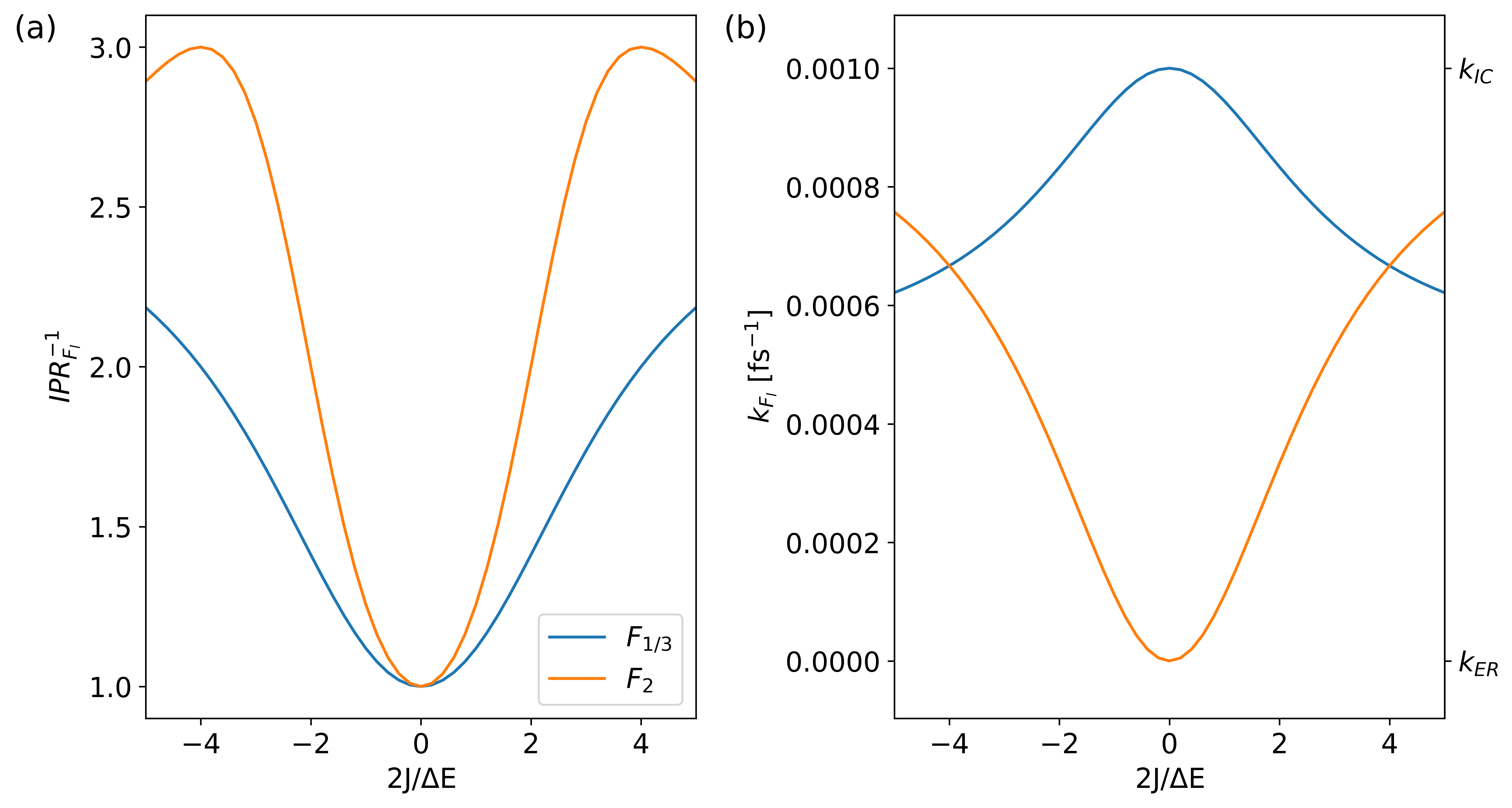}
    \caption{(a) The reciprocal of the Inverse Partecipation Ratio (IPR) for the eigenstates of the two-exciton eigenstates $\ket{F_{2}}$ (orange), $\ket{F_{1}}$ and $\ket{F_{3}}$ (blue, overlapped) as a function of $2J / \Delta E$. (b) Total rates from the  two- to one-exciton manifold $k_{F_{2}}$ (orange), $k_{F_{2}}$ and $k_{F_{3}}$ (blue, overlapped) as a function of $2J / \Delta E$. For reference, the limiting values of these rates, namely the Internal Conversion (IC) and Exciton Recombination (ER) rates, are reported.}
    \label{fig:rates_IPR}
\end{figure}

\section{Fourth-order response functions}

The fourth-order response functions contributing to the rephasing signal in A-2DES are:
\begin{equation}
\begin{split}
R_{\text{GSB}}(t_{d}, t_3, t_2, t_1) &= - \left(\frac{i}{\hbar} \right)^{4} \Tr\left\{\hat{S} \mathcal{G}(t_{d}) \mathcal{M}^{-}_{B} \mathcal{G}(t_{3}) \mathcal{M}^{+}_{K} \mathcal{G}(t_{2}) \mathcal{M}^{+}_{B} \mathcal{G}(t_{1}) \mathcal{M}^{-}_{B}  \rho(0) \right\} \\
R_{\text{SE}}(t_{d}, t_3, t_2, t_1) &= - \left(\frac{i}{\hbar} \right)^{4} \Tr\left\{ \hat{S} \mathcal{G}(t_{d}) \mathcal{M}^{-}_{B} \mathcal{G}(t_{3}) \mathcal{M}^{+}_{B} \mathcal{G}(t_{2}) \mathcal{M}^{+}_{K} \mathcal{G}(t_{1}) \mathcal{M}^{-}_{B}  \rho(0) \right\} \\
R_{\text{ESAI}}(t_{d}, t_3, t_2, t_1) &= - \left(\frac{i}{\hbar} \right)^{4} \Tr\left\{ \hat{S} \mathcal{G}(t_{d}) \mathcal{M}^{-}_{K} \mathcal{G}(t_{3}) \mathcal{M}^{+}_{K} \mathcal{G}(t_{2}) \mathcal{M}^{+}_{K} \mathcal{G}(t_{1}) \mathcal{M}^{-}_{B}  \rho(0) \right\} \\
R_{\text{ESAII}}(t_{d}, t_3, t_2, t_1) &= + \left(\frac{i}{\hbar} \right)^{4} \Tr\left\{ \hat{S} \mathcal{G}(t_{d}) \mathcal{M}^{-}_{B} \mathcal{G}(t_{3}) \mathcal{M}^{+}_{K} \mathcal{G}(t_{2}) \mathcal{M}^{+}_{K} \mathcal{G}(t_{1}) \mathcal{M}^{-}_{B}  \rho(0) \right\} \\
\end{split}
\label{eq:response functions liouville}
\end{equation}
where, for the superoperator $\mathcal{M}$, the superscript $+$ ($-$) represent the rotating (counter-rotating) component of the dipole moment operator, while the subscript $K$ ($B$) indicate if it acts from the left (right) of the density matrix.

\begin{figure}[H]
    \centering
    \includegraphics[width=\linewidth]{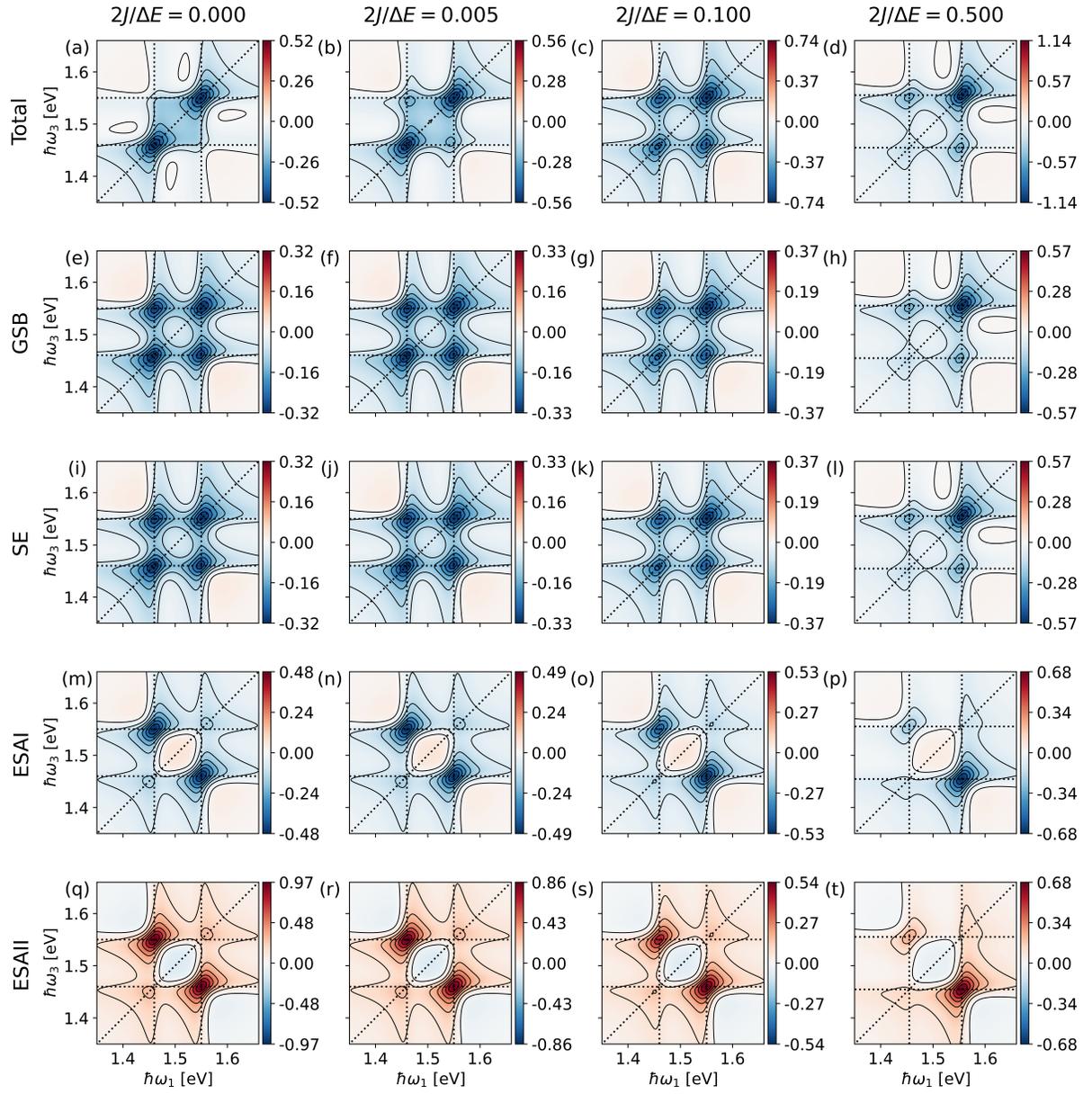}
    \caption{(a-d) Total rephasing, (e-h) GSB, (i-l) SE, (m-p) ESAI and (q-t) ESAII spectra at waiting-time $t_2 = 0\ \si{fs}$ simulated at different values of $2J / \Delta E = 0.000$, $0.005$, $0.100$, $0.500$.}
    \label{fig:components_spectra}
\end{figure}

\begin{figure}[H]
    \centering
    \includegraphics[width=\linewidth]{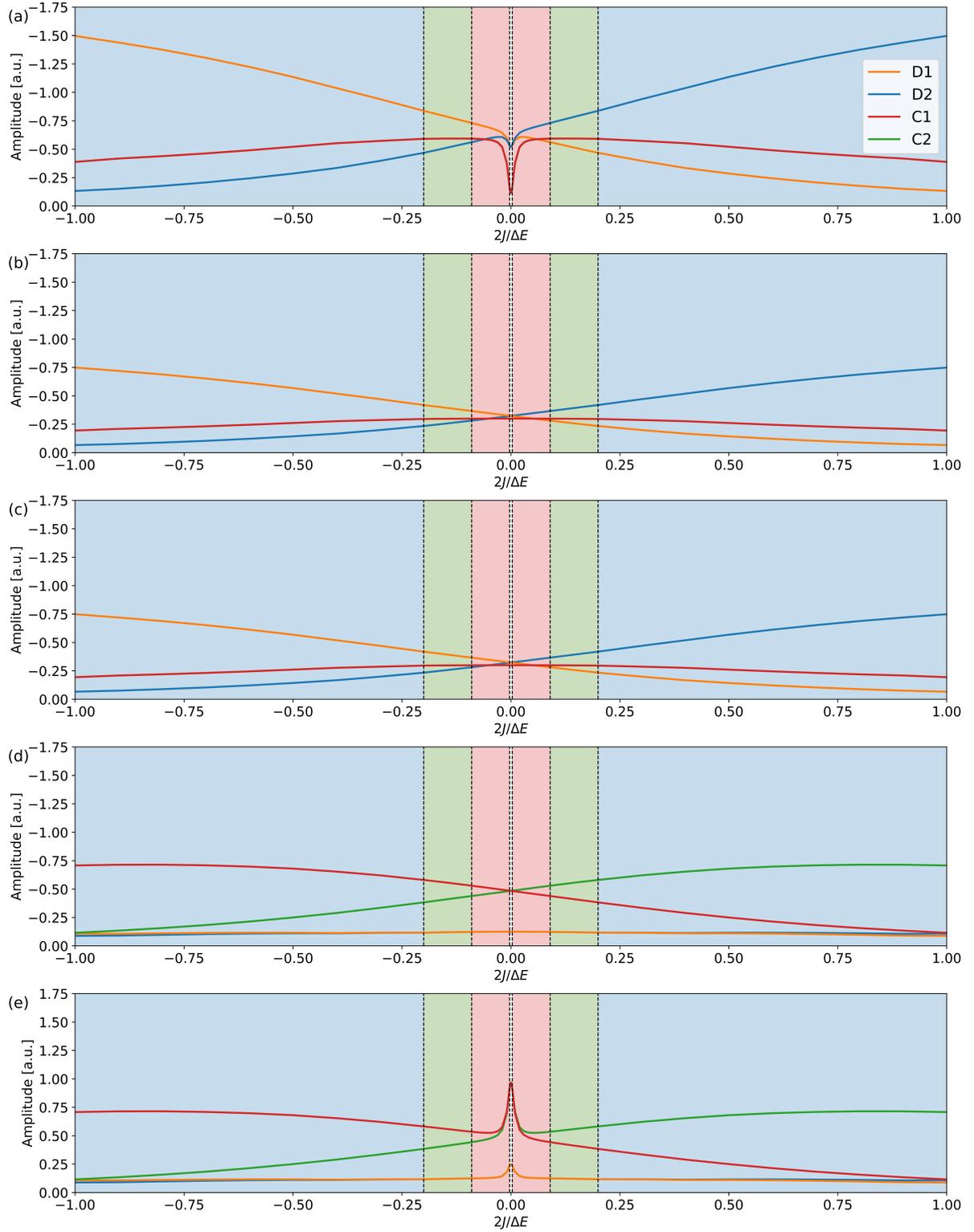}
    \caption{(a) Amplitude of total rephasing, (b) GSB, (c) SE, (d) ESAI and (e) ESAII spectral features as a function of $2J / \Delta E$ at $t_2$ = 0 fs. The regions corresponds to: (white area) no cross peaks, (red area) incoherent mixing, (green area) ESA-free, (blue area) delocalization.}
    \label{fig:profile_components}
\end{figure}

If we neglect the non-secular contributions during the delay times $t_{1}$ and $t_{3}$ and we consider waiting time $t_{2} = 0\ \si{fs}$, these pathways are proportional to: 
\begin{equation}
\begin{split}
R_{\text{GSB}}(t_{d}, t_3, t_2, t_1) &\propto \sum_{k k'} \mu_{E_{k'}G} \mu_{E_{k'}G} \mu_{E_{k}G} \mu_{E_{k}G} \\
R_{\text{SE}}(t_{d}, t_3, t_2, t_1) &\propto \sum_{k k'} \mu_{E_{k'}G} \mu_{E_{k}G} \mu_{E_{k'}G} \mu_{E_{k}G} \\
R_{\text{ESAI}}(t_{d}, t_3, t_2, t_1) &\propto \sum_{k k'l} \mu_{F_{l}E_{k}} \mu_{F_{l}E_{k'}} \mu_{E_{k'}G} \mu_{E_{k}G} \\
R_{\text{ESAII}}(t_{d}, t_3, t_2, t_1) &\propto \sum_{k k'l} \mu_{F_{l}E_{k}} \mu_{F_{l}E_{k'}} \mu_{E_{k'}G} \mu_{E_{k}G} \\
\end{split}
\label{eq:response functions hilbert}
\end{equation}
In Figs. \ref{fig:components_spectra} and \ref{fig:profile_components}, we report the spectra and amplitude, respectively, for the total rephasing and for each class of pathways for different values of excitonic coupling.

\section{Analysis of the peak amplitude}

From the fourth-order response functions in Eq. \ref{eq:response functions hilbert}, each contribution to the spectrum is weighted by the product of four transition dipole moments. The possible combinations of indices in the eigenstate basis for each class of pathways are reported in Table \ref{tab:dipole moments}.

\begin{table}[H]
\centering
\label{tab:dipole moments}
\caption{Combinations of indices for the amplitude of diagonal and cross peaks in GSB, SE, ESAI and ESAII pathways. The indices takes values $k, k' = 1, 2$ and $l = 1, 2, 3$.}
\begin{tabular}{l l l l}
\hline
\hline
Pathway & Peak Type & Oscillating along $t_2$ & Indices Combinations\\
\hline
GSB/SE & Diagonal & No & $k = k'$ \\
GSB & Cross & No & $k \neq k'$ \\
SE & Cross & Yes & $k \neq k'$ \\
ESAI/ESAII & Cross & No & $k = k', l$ \\
ESAI/ESAII & Cross & Yes & $k \neq k', l$ \\
\hline
\hline
\end{tabular}
\end{table}
 
Using Eq. \ref{eq:dipole_moments_1} and the definitions in Table \ref{tab:diagonalization}, we consider now how the different contributions behave in the weak-coupling regime, by using Taylor expansion.
For the GSB and SE pathways, truncated to the first order, the amplitude of diagonal peaks has a linear dependence on $J$:
\begin{equation}
\begin{split}
|\mu_{E_{1}G}|^4 &= |\mu_{e_{1}g_{1}}|^4 - \frac{4 J}{\Delta E} \mu_{e_{1}g_{1}}^3 \mu_{e_{2}g_{2}} \\
|\mu_{E_{2}G}|^4 &= |\mu_{e_{2}g_{2}}|^4 + \frac{4 J}{\Delta E} \mu_{e_{1}g_{1}} \mu_{e_{2}g_{2}}^3
\end{split}
\label{eq:dipole diagonal weak}
\end{equation}
while for cross peaks, expanding up to the second order, is:
\begin{equation}
\begin{split}
|\mu_{E_{1}G}|^2 |\mu_{E_{2}G}|^2 = |\mu_{e_{1}g_{1}}|^2 |\mu_{e_{2}g_{2}}|^2 + J (\mu_{e_{1}g_{1}}^3 \mu_{e_{2}g_{2}} - \mu_{e_{1}g_{1}} \mu_{e_{2}g_{2}}^3)
+ \frac{J^2}{\Delta E^2} (|\mu_{e1g1}|^4 + |\mu_{e2g2}|^4 - 6|\mu_{e1g1}|^2 |\mu_{e2g2}|^2)
\end{split}
\label{eq:dipole cross GSB-SE weak}
\end{equation}
whose linear dependence on $J$ vanished if the transition dipole moment of the chromophores are equal, as assumed in this work. As a result, cross peaks of GSB and SE are insensitive to the excitonic coupling in this regime (i.e., green region in Figure 3). Notice that this is also the amplitude of the pathways oscillating during the waiting time $t_{2}$.

Using the same approach, truncating the Taylor expansion at the second order, we derived expressions for cross-peaks amplitude for ESAI and ESAII, both for the non-oscillating pathways:
\begin{equation}
\begin{split}
|\mu_{E_{1}G}|^2 |\mu_{F_{1/3}E_{1}}|^2 &= \frac{\alpha^2 J^2}{\Delta E^2} |\mu_{e_{1}g}|^2 |\mu_{e_{2}g}|^2 \\
|\mu_{E_{2}G}|^2 |\mu_{F_{1/3}E_{2}}|^2 &= \frac{\alpha^2 J^2}{\Delta E^2} |\mu_{e_{1}g}|^2 |\mu_{e_{2}g}|^2 \\
|\mu_{E_{1}G}|^2 |\mu_{F_{2}E_{1}}|^2 &= |\mu_{e_{1}g_{1}}|^2 |\mu_{e_{2}g_{2}}|^2 - \frac{J^2}{\Delta E^2} (|\mu_{e1g1}|^4 + |\mu_{e2g2}|^4 + 2(\alpha^2 + 1)|\mu_{e1g1}|^2 |\mu_{e2g2}|^2)
\\
|\mu_{E_{2}G}|^2 |\mu_{F_{2}E_{2}}|^2 &= |\mu_{e_{1}g_{1}}|^2 |\mu_{e_{2}g_{2}}|^2 - \frac{J^2}{\Delta E^2} (|\mu_{e1g1}|^4 + |\mu_{e2g2}|^4 + 2(\alpha^2 + 1)|\mu_{e1g1}|^2 |\mu_{e2g2}|^2)
\\
\end{split}
\end{equation}
and the oscillating pathways:
\begin{equation}
\begin{split}
\mu_{E_{1}G} \mu_{E_{2}G} \mu_{F_{1/3}E_{1}} \mu_{F_{1/3}E_{2}} &= \frac{\alpha^2 J^2}{\Delta E^2} |\mu_{e_{1}g}|^2 |\mu_{e_{2}g}|^2 \\
\mu_{E_{1}G} \mu_{E_{2}G} \mu_{F_{2}E_{1}} \mu_{F_{2}E_{2}} &= |\mu_{e_{1}g_{1}}|^2 |\mu_{e_{2}g_{2}}|^2
\end{split}
\end{equation}
At weak excitonic coupling, only $|\mu_{E_{1}G}|^2 |\mu_{F_{2}E_{1}}|^2$, $|\mu_{E_{2}G}|^2 |\mu_{F_{2}E_{2}}|^2$ and $\mu_{E_{1}G} \mu_{E_{2}G} \mu_{F_{2}E_{1}} \mu_{F_{2}E_{2}}$ give rise to cross peaks in the spectrum.

In A-2DES, the amplitude of coherent oscillations during $t_{2}$ is directly related to that of SE pathways, assuming an ESA-free scenario.
Instead, in C-2DES, the amplitude of coherent oscillations is given by the difference between SE and ESA contributions (Table \ref{tab:dipole moments}). For SE, the amplitude is $|\mu_{E_{1}G}|^2 |\mu_{E_{2}G}|^2$. For ESA, three different indices combination should be considered. However, at lower excitonic coupling only the amplitude $\mu_{E_{1}G} \mu_{E_{2}G} \mu_{F_{2}E_{1}} \mu_{F_{2}E_{2}}$ is different from zero. Therefore, it is sufficient to calculate only one term, expanding it up to the second order:
\begin{equation}
\begin{split}
|\mu_{E1G}|^2 |\mu_{E2G}|^2 - \mu_{E1G} \mu_{E2G} \mu_{F2E1} \mu_{F2E2} =& J (\mu_{e1g1}^3 \mu_{e2g2} - \mu_{e1g1} \mu_{e2g2}^3) \\ &+ \frac{2J^2}{\Delta E^2} (|\mu_{e1g1}|^4 + |\mu_{e2g2}|^4 - (2- \alpha ^2)|\mu_{e1g1}|^2 |\mu_{e2g2}|^2)
\end{split}
\end{equation}
Assuming the site-basis dipole moments to be equal, the amplitude of the oscillations in the C-2DES is null at weak excitonic coupling and quadratically dependent from $J$ at larger values, assuming a non-zero excitonic coupling in the two-exciton manifold ($\alpha \ne 0$). Comparing this result with the one related to A-2DES in Eq. \ref{eq:dipole cross GSB-SE weak}, we conclude that in the weak-coupling regime, coherent oscillations are expected only in A-2DES and not in C-2DES.

\section{Fourier power spectra}

\begin{figure}[H]
    \centering
    \includegraphics[width=\linewidth]{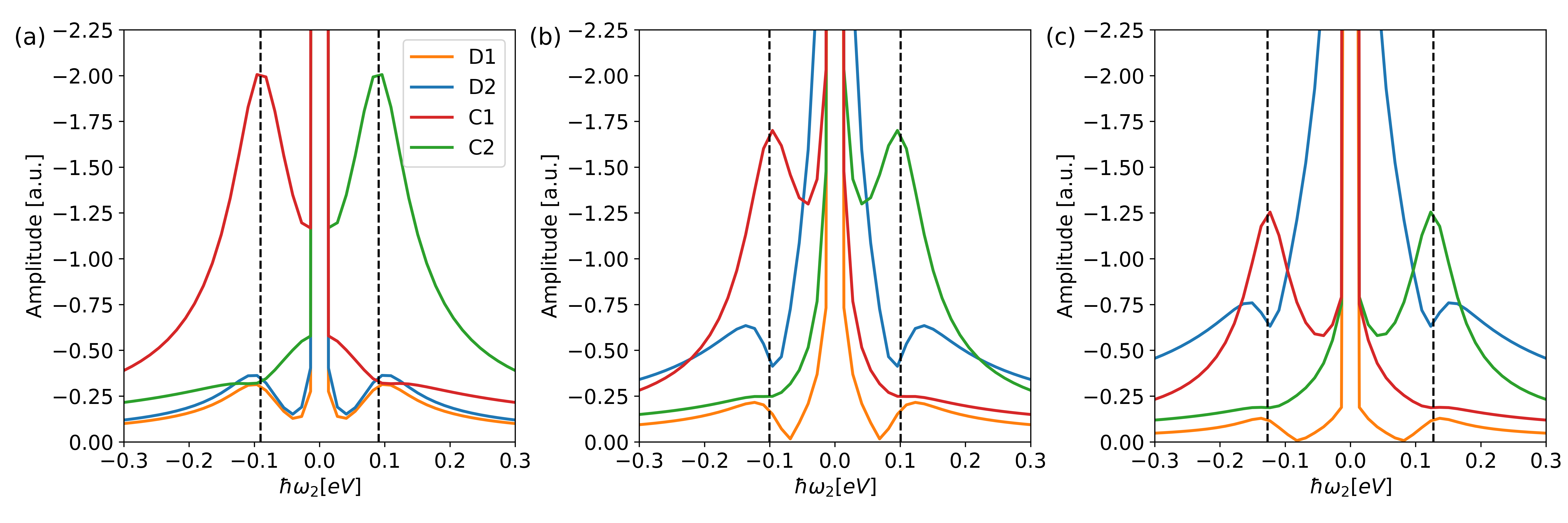}
    \caption{Absolute value of the power spectrum along $\hbar \omega_{2}$ for different values of $2J/\Delta E = 0.1$, $0.5$, $1.0$. Dashed lines highlight the energy splitting between one-exciton eigenstates.}
    \label{fig:t2_fft}
\end{figure}

Fourier power spectra were calculated by taking the Fourier transform the signal along the waiting time $t_2$. As can be seen in Fig. \ref{fig:t2_fft}, the spectral features exhibit frequency components compatible with the energy splitting between states in the one-exciton manifold.

\section{Non-perturbative simulation}

The non-perturbative simulations are obtained using the phase-modulation scheme detailed in Refs. \cite{bruschi:2022}.

\begin{figure}[H]
    \centering
    \includegraphics[width=0.9\linewidth]{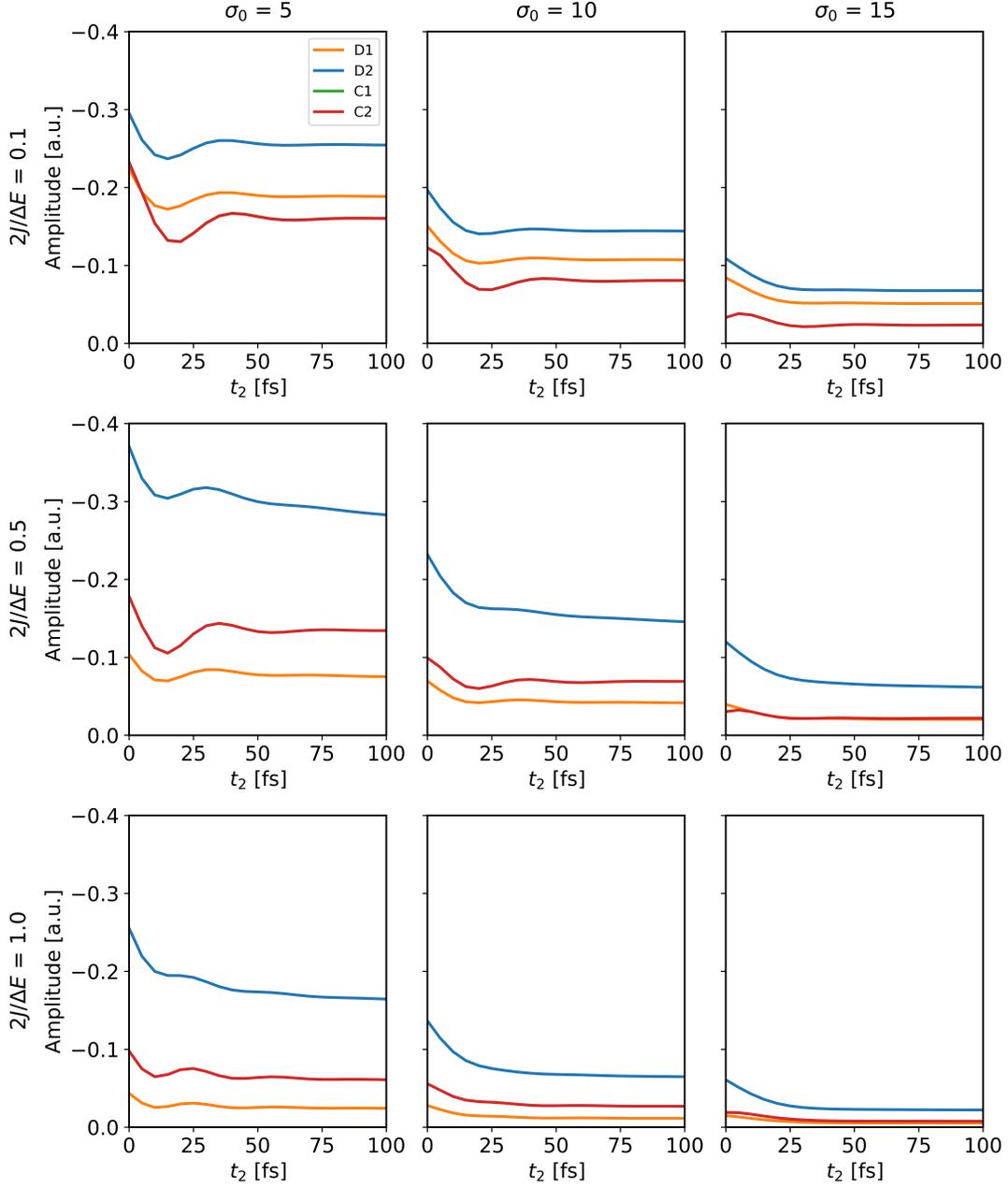}
    \caption{Amplitudes of spectral features in A-2DES as a function of the waiting time $t_{2}$ for different values of pulse duration $\sigma_{0}$ and excitonic coupling $2J/\Delta E$.}
    \label{fig:non_perturbative}
\end{figure}

The strength of the light-matter interaction is $\mu_{n} E_{0} = 0.008\ \si{eV}$, while the energy of the pulses is set to $\hbar\omega_{0} = 1.505\ \si{eV}$. The modulation frequencies used are $f_{1} = 0\ \si{Hz}$, $f_{2} = 250\ \si{Hz}$, $f_{3} = 600\ \si{Hz}$ and $f_{4} = 1000\ \si{Hz}$, with an inter-train delay time of $3000\ si{s}$, corresponding to $179$ repetitions of phase modulation.
The delay times $t_{1}$ and $t_{3}$ are scanned from $0$ to $300\ \si{fs}$ in steps of $10\ \si{fs}$, while the waiting time is scanned from $0$ to $100\ \si{fs}$ in steps of $5\ \si{fs}$. The data are acquired in a fully-rotating frame.
In Fig. \ref{fig:non_perturbative}, we report the coherent dynamics during the waiting time for three different pulse duration corresponding to standard deviation $\sigma_0 = 5$, $10$, $15$ considering different values of $2J/\Delta E$.

\printbibliography